\documentclass[12pt,a4paper,final]{iopart}

\usepackage{iopams}  
\usepackage{graphicx}
\usepackage[breaklinks=true,colorlinks=true,linkcolor=blue,urlcolor=blue,citecolor=blue]{hyperref}
\expandafter\let\csname equation*\endcsname\relax
\expandafter\let\csname endequation*\endcsname\relax
\usepackage{amsmath}
\usepackage{gensymb}
\usepackage[dvipsnames]{xcolor}
\usepackage{textcomp}
\usepackage{multirow}
\usepackage{commath}
\usepackage{siunitx}
\usepackage{url}
\usepackage{doi}
\usepackage{makecell}
\usepackage{tabularx}

\usepackage{array}
\newcommand{\PreserveBackslash}[1]{\let\temp=\\#1\let\\=\temp}
\newcolumntype{C}[1]{>{\PreserveBackslash\centering}p{#1}}
\newcolumntype{R}[1]{>{\PreserveBackslash\raggedleft}p{#1}}
\newcolumntype{L}[1]{>{\PreserveBackslash\raggedright}p{#1}}
\makeatletter
\let\@fnsymbol\@arabic
\renewcommand{\@makefnmark}{\makebox{\normalfont$^{\@thefnmark}$}}
\makeatother
\begin{document}
\title[]{Fiducial displacements with improved accuracy for the global network of gravitational wave detectors}
\author{D~Bhattacharjee$^{1,2}$, Y~Lecoeuche$^1$, S~Karki$^{1,2}$, J~Betzwieser$^3$, V~Bossilkov$^{1,4}$, S~Kandhasamy$^5$,  E~Payne$^{1,6}$ and R~L~Savage$^1$}
\address{$^1$ LIGO Hanford Observatory, Richland, WA 99352 USA}
\address{$^2$ Institute of Multi-messenger Astrophysics and Cosmology, Missouri University of Science and Technology, Rolla, MO 65409 USA }
\address{$^3$ LIGO Livingston Observatory, Livingston, LA 70754 USA}
\address{$^4$ The University of Western Australia, Crawley WA 60069, Australia}
\address{$^5$ Inter-University Centre for Astronomy and Astrophysics, Pune-411007, India}
\address{$^6$ OzGrav, School of Physics \& Astronomy, Monash University, Clayton  VIC 3800, Australia}
\ead{rsavage@caltech.edu}
\date{March 14, 2020}
\setcounter{footnote}{6}
\begin{abstract}
As sensitivities improve and more detectors are added to the global network of gravitational wave observatories, calibration accuracy and precision are becoming increasingly important.  Photon calibrators, relying on power-modulated auxiliary laser beams reflecting from suspended interferometer optics, enable continuous calibration by generating  displacement fiducials proportional to the modulated laser power.    Developments in the propagation of laser power calibration via transfer standards to on-line power sensors monitoring the modulated laser power have enabled generation of length fiducials with improved accuracy.  Estimated uncertainties are almost a factor of two smaller than the lowest values previously reported.  This is partly due to improvements in methodology that have increased confidence in the results reported.  Referencing the laser power calibration standards for each observatory to a single transfer standard enables reducing relative calibration errors between elements of the detector network. Efforts within the national metrology institute community to realize improved laser power sensor calibration accuracy are ongoing.
\end{abstract}
\submitto{\CQG}
\section{Introduction}
\label{sec:intro}
Since the first direct detection of gravitational waves (GWs) from a coalescing binary black hole system in 2015\,\cite{detection}, the rate of detections by the LIGO\footnote{Laser Interferometer Gravitational-wave Observatory}~\cite{detectorpaper} and Virgo\,\cite{VirgoPaper} observatories has increased\,\cite{O2Catalog}, yielding insight into properties, formation processes, and populations estimates of GW sources.
As the Japanese KAGRA\,\cite{kagradet} detector comes on-line and is eventually joined by a third LIGO detector located in India\,\cite{LIGOIndia}, and as the sensitivities of the currently-operating detectors continue to improve,  the rate of detections by the global network is expected to increase to  several per day\,\cite{detectionrate}.
To maximize the scientific benefit of these detections, accurate and precise calibration of the detectors is essential. In 2009, L.~Lindblom estimated that calibration accuracy of 0.5\,\% or better would be required to optimally extract the information encoded in the signals\,\cite{Lindblom}. Subsequent analyses have also highlighted the importance of reducing calibration uncertainties\,\cite{VitaleCal, GW150914Cal}.  Accurate determination of the distance to GW sources requires low overall network calibration uncertainty.  Also, the  relative calibration accuracy between detectors in the global network plays an important role in sky localization of sources, enabling improved sky maps for follow-up observations by electromagnetic observatories\,\cite{relcalibimp2}.

In 2017, the LIGO and Virgo collaborations, together with over seventy electromagnetic observatories, reported the multi-messenger observation of a binary neutron star inspiral\,\cite{GW170817}.  These observations enabled an independent measurement of the Hubble parameter\,\cite{hubble}, albeit with insufficient precision to resolve the tension between the results reported by the {\em{Planck}}\,\cite{planck} and the {\em {SHoES}}\,\cite{shoes} collaborations. This result was constrained by the signal-to-noise ratio (SNR) rather than calibration accuracy.   As the number and SNR of GW detections increase, measuring the Hubble parameter with the $\sim$\,1\,\% accuracy that is needed to resolve the tension in the current estimates is expected within the next decade.  It will require detector calibration accuracy of 1\,\% or better\,\cite{hubbleat1per}.    

\begin{figure}[t]%
    \begin{center}
    \includegraphics[trim= 0.2cm 0.2cm 0.2cm 0.2cm, clip=true, width = 0.9\textwidth]{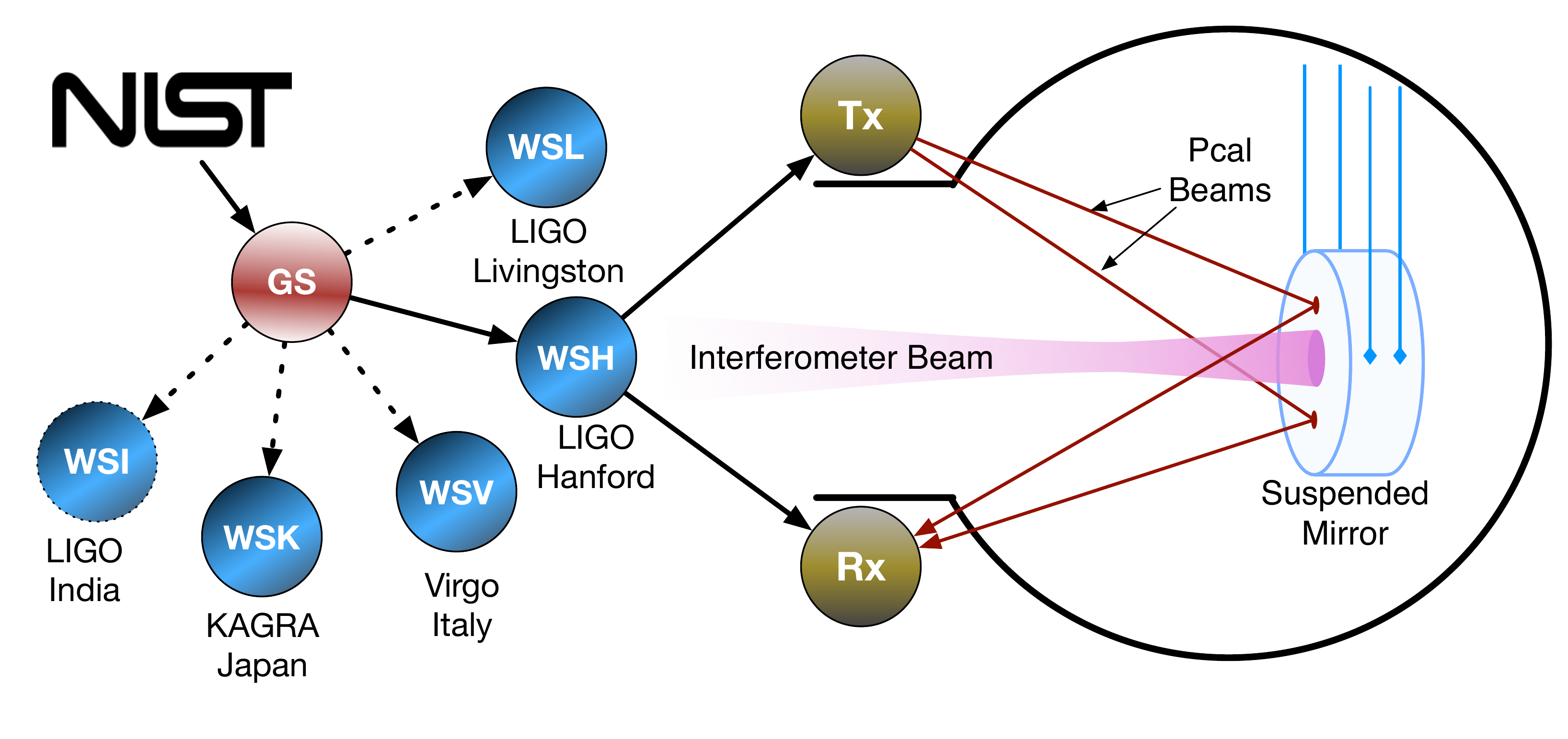}
    \caption{Schematic diagram showing the transfer of laser power calibration from SI units via calibration of a {\em Gold} transfer standard (GS) by NIST.   Then from the GS to {\em Working}  transfer standards (WS), one for each observatory, and then to the power sensors (Tx and Rx) located at the interferometer end stations. These calibrated power sensors enable on-line monitoring of the amplitudes of the fiducial periodic displacements induced by the power modulated Pcal beams reflecting from the suspended mirror.}
    \label{fig:calib_transfer}
    \end{center}
\end{figure}

Current gravitational-wave interferometers use systems that are referred  to as {\em photon calibrators} (Pcals) to produce periodic  fiducial displacements of suspended interferometer mirrors via photon radiation pressure\,\cite{RSIpaper,VirgoPcal,InoueGcal,GEOPcal}.  A schematic diagram of a Pcal system installed at an end station of one of the LIGO interferometers is shown in \fref{fig:calib_transfer}. 
Power-modulated auxiliary laser beams reflecting from the interferometer mirrors cause differential relative arm length variations that mimic the variations induced by gravitational waves.  The magnitudes of the induced length variations are proportional to the amplitudes of the modulated laser power.  Thus the interferometer displacement accuracy depends directly on the accuracy of the calibration of the Pcal laser power sensors.  The scheme for transferring laser power calibration from the US National Institute of Standards and Technology (NIST) to the Pcal power sensors, and for coordinating the relative calibration of all of the detectors in the global GW network, is also shown schematically in \fref{fig:calib_transfer}.

\subsection{Pcal forces and induced displacements}
\label{subsec:IntroForces}
\begin{figure}[t]%
    \begin{center}
    \includegraphics[trim=0cm 0.5cm 0cm 0.5cm, clip=true, width = 1.0\textwidth]{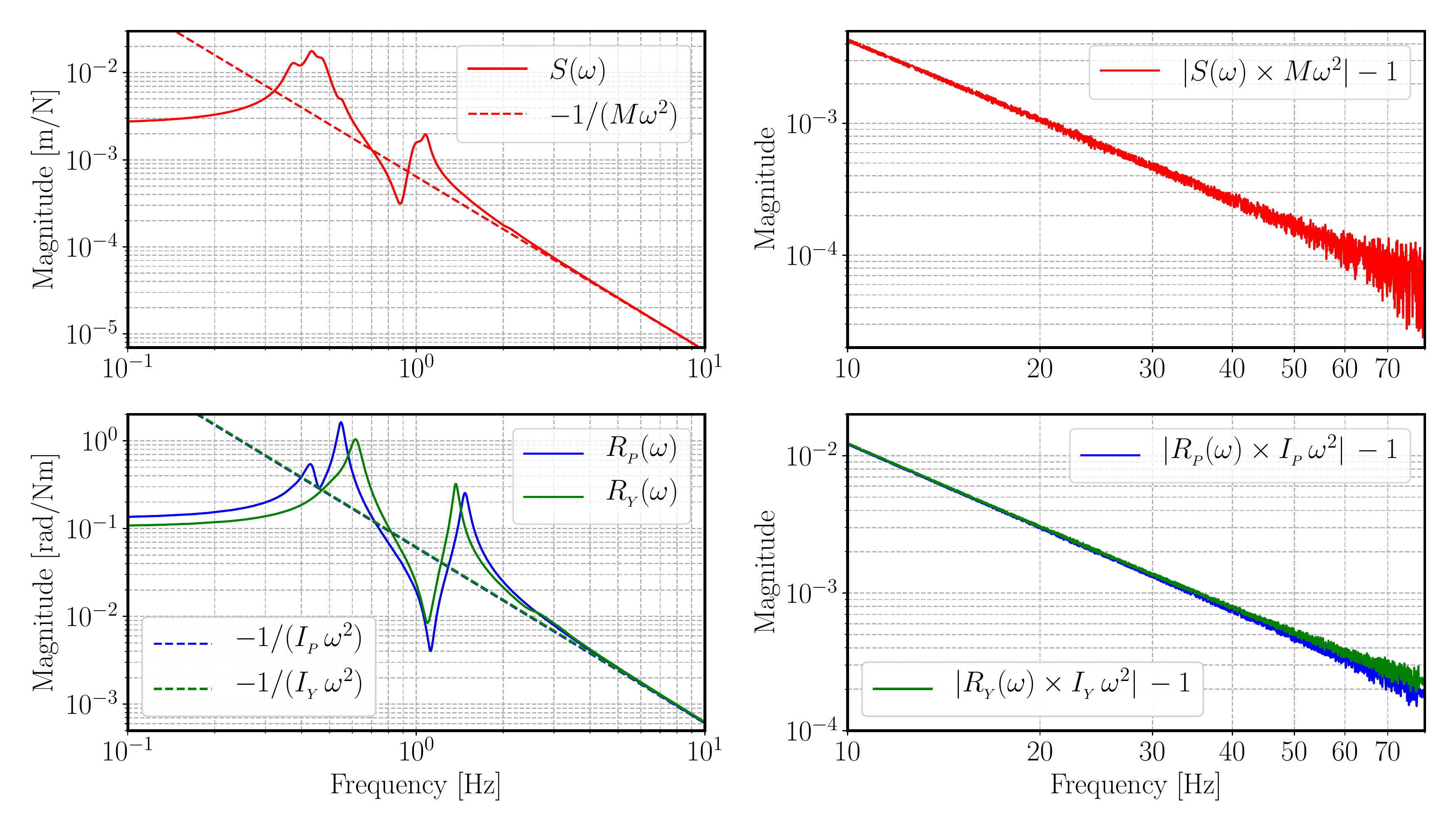}
    \caption{{\em Upper-left panel}: Modeled force-to-displacement transfer function for a suspended LIGO end test mass and for a free mass, -$1/(M \omega^2)$.  {\em Lower-left panel}: Modeled torque-to-rotation transfer functions for a suspended LIGO end test mass for both pitch and yaw, and for a free mass, -$1/(I \omega^2$). {\em Upper-right} and {\em Lower-right panels}: Discrepancy between the modeled and free-mass transfer functions. Above 20~Hz, the $S(\omega)$ discrepancy is less than 0.1\,\%, and $R(\omega)$ discrepancy is less than 0.3\,\%.}%
    \label{fig:TF}
    \end{center}
\end{figure}

When power-modulated Pcal laser beams reflect from the surface of the mirror suspended at the end of one of the arms of a gravitational wave interferometer, referred to as an {\em end test mass} (ETM), the periodic force exerted on the optic is given by 
\begin{equation}
F(\omega)  = \frac{2 \cos\theta}{c} P(\omega) \, ,
\label{eq:force}
\end{equation}
where $\theta$ is the angle of incidence of the Pcal beams on the ETM surface, $c$ is the speed of light, P($\omega$) is the amplitude of the modulated laser power reflected from the test mass, and  $\omega = 2\pi f$ is the angular frequency of the power modulation. 
Conversion of the periodic Pcal-induced forces to displacements requires the force-to-length transfer function of the suspended mirror, $S(\omega)$ in units of m/N. $S(\omega)$ for a LIGO 40\,kg ETM is plotted in the upper-left panel of \fref{fig:TF} together with the transfer function for a free mass, $S(\omega) = -1/(M\omega^2)$, where $M$ is the mass of the suspended optic.
As shown in the upper-right panel of \fref{fig:TF}, at frequencies above 20\,Hz $S(\omega)$ is well approximated (within 0.1\,\%) by the response of a free-mass.

The Pcal forces can also induce unintended rotation of the test mass due to power imbalance between the two Pcal beams\footnote{LIGO uses a two-beam configuration to minimize local elastic deformation of the ETM surface in the region sensed by the interferometer beam\,\cite{P080118,LocalDeformation}} or beam positions that are offset from their nominal locations\footnote{The nominal locations for the two Pcal beams are diametrically opposed and displaced above and below the ETM center by 111.6~mm.  These locations are close to the nodal circle of the {\em drumhead} natural deformation mode of the ETM and are chosen to minimize sensing of {\em bulk} elastic deformation of the ETM by the interferometer\,\cite{P1900127}.}. If the interferometer beam is also offset from its nominal location at the center of the ETM surface, the rotations will be sensed by the interferometer as length variations. Conversion of the unintended rotation to length variation depends on the torque-to-rotation transfer function for the suspended optic, $R(\omega)$, in units of 1/(N\,m), and the dot product between the displacement vector for the center of force of the Pcal beams, $\vec{a}$, and the displacement vector for the interferometer beam, $\vec{b}$, as given in \eref{eq:pcaldisp}.  $R(\omega)$ is plotted in the lower-left panel of \fref{fig:TF} for a LIGO ETM for both pitch and yaw rotations.   As shown in the lower-right panel  of \fref{fig:TF}, at frequencies above 20\,Hz, like $S(\omega)$, $R(\omega)$ is well approximated (within 0.3\,\%) by the response of a free-mass, $R(\omega)\simeq-1/(I \omega^2)$, where $I$ is the moment of inertia of the suspended optic about the relevant axis of rotation.

Taking into account the longitudinal displacement as well as the apparent displacements caused by unintended rotations\,\cite{P080118}, the fiducial length modulation induced by Pcal forces, $x(\omega)$, is given by
\begin{equation}
\label{eq:pcaldisp}
\begin{split}
x(\omega) & = \frac{2 \cos\theta}{c}P(\omega)   \left[S(\omega)+ R(\omega)(\vec{a}\cdot\vec{b})\right] \simeq - \frac{2 \cos\theta}{M c \, \omega^2 }P(\omega) \left[1 + \frac{M}{I} (\vec{a}\cdot\vec{b})\right] .
\end{split}
\end{equation}
 In practice,  we do not know the magnitude or direction of the Pcal center of force displacement vector, $\vec{a}$; we can only estimate the maximum magnitude, as described in \sref{subsec:ResultsForcecoeff}. Therefore the second term in the square brackets determines the relative uncertainty in the displacement amplitude introduced by the unintended rotation of the ETM , which we define to be $\epsilon_{rot}$.
 As \eref{eq:pcaldisp} shows, calculation of the displacements induced by the Pcal systems requires estimates of three parameters: the angle of incidence of the Pcal beams on the ETM surface, the mass of the ETM, and the laser power reflecting from the highly-reflective ETM surface inside the vacuum envelope.  Accurate measurement of these parameters, especially $P(\omega)$, in order to accurately estimate the Pcal displacements is the focus of this paper.
 
 It is organized as follows:  in \sref{sec:method} we discuss methodology for calibrating Pcal power sensors that are located outside the vacuum envelope in terms of the power reflecting from the ETM. Improvements in measurement methods and compensation for temperature variations are also presented.  We also discuss a new method for comparing displacement factors from both end stations to calculate {\em combined} displacement factors with reduced uncertainty.  In \sref{sec:results} we use measurements made with the LIGO Hanford Observatory (LHO) interferometer during the O3 observing run, from April 2019 to March 2020, to demonstrate the application of the methods described in \sref{sec:method}.  We also give detailed descriptions of how uncertainties in the various measured and calculated parameters are estimated.  In \sref{sec:Conc} we summarize the results presented in this paper and discuss prospects for further improvements in the accuracy and uncertainty of fiducial displacements generated to calibrate gravitational wave detectors.  A list of the symbols used in this paper and the parameters they represent is included  in \tref{tab:allParams} in Appendix~A.
\section{Methodology}
\label{sec:method}
\subsection{Calibration of the power sensors}
\label{subsec:MethodSensorcal}
The LIGO Pcal systems incorporate a laser power sensor that is located outside the vacuum envelope and receives almost all of the laser power exiting the vacuum chamber after reflecting from the ETM (the {\em Rx} sensor in \fref{fig:calib_transfer}).  A second sensor  ({\em Tx}) samples a small fraction of the laser power directed into the vacuum chamber and is used for power calibration and optical efficiency measurements.  Calibrating the Rx sensor in terms of power reflected from the ETM requires compensating for optical losses between the ETM and the sensor.

The power reflectivity of the ETM is greater than 0.9999 but the anti-reflection coated vacuum windows and the relay mirrors located inside the vacuum envelope\,\cite{RSIpaper} reduce the optical efficiency, $\eta$, between the transmitter and receiver modules to approximately 0.985 - 0.990, i.e.\ the overall optical loss is about \num{1.0} to \SI{1.5}{\percent}. 
Measurements made inside the vacuum envelope when the system is vented to atmosphere enable apportioning the overall optical efficiency between the input side (between the Tx sensor and the ETM), $\eta_{_T}$, and the output side (between the ETM and the Rx sensor), $\eta_{_R}$.  The measured efficiency ratio, $\beta = \eta_{_{T}} / \eta_{_{R}}$, together with the overall optical efficiency that can be measured outside the vacuum envelope, enables calculation of the optical efficiency factors as
\begin{equation}
\label{eq:etaRetaT}
  \eta_{_{R}} = \sqrt{\eta/\beta} \ {\rm and} \ \eta_{_{T}} = \sqrt{\eta \, \beta}.  
\end{equation}
The power reflecting from the ETM is thus estimated by

\begin{equation}
\label{eq:power_prime}
    P(\omega) = P_{_R}(\omega) / \eta_{_R}
\end{equation}
where $P_{_{R}}(\omega)$ is the power measured at the Rx sensor.  The power at the ETM can also be calculated as $ P(\omega) = \eta_{_T} \,P_{_T}(\omega)$ with $P_{_{T}}(\omega)$ the power measured on the Tx side.  However, this sensor is less reliable because it is subject to variations in the beamsplitter that reflects the small sample of the input light and because it is insensitive to changes in the optical efficiency, $\eta$.

Calibration of the power sensors at the interferometer end stations is realized by a three-step process as shown schematically in \fref{fig:calib_transfer}: $i$) a transfer standard referred to as the {\em Gold Standard} (GS) is calibrated to SI units at NIST in Boulder, Colorado; $ii$) the calibration of the GS is propagated to additional transfer standards referred to as {\em Working Standards} (WS), one for each observatory, by making responsivity ratio measurements in a dedicated laboratory setup at the LHO, and $iii$) the calibration of a WS is transferred to end station power sensors at each observatory by making responsivity ratio measurements at the end stations.

\begin{figure}[t]%
    \begin{center}
    \includegraphics[trim= 0.2cm 0.8cm 1.2cm 1.9cm, clip=true, width = 1.0\textwidth]{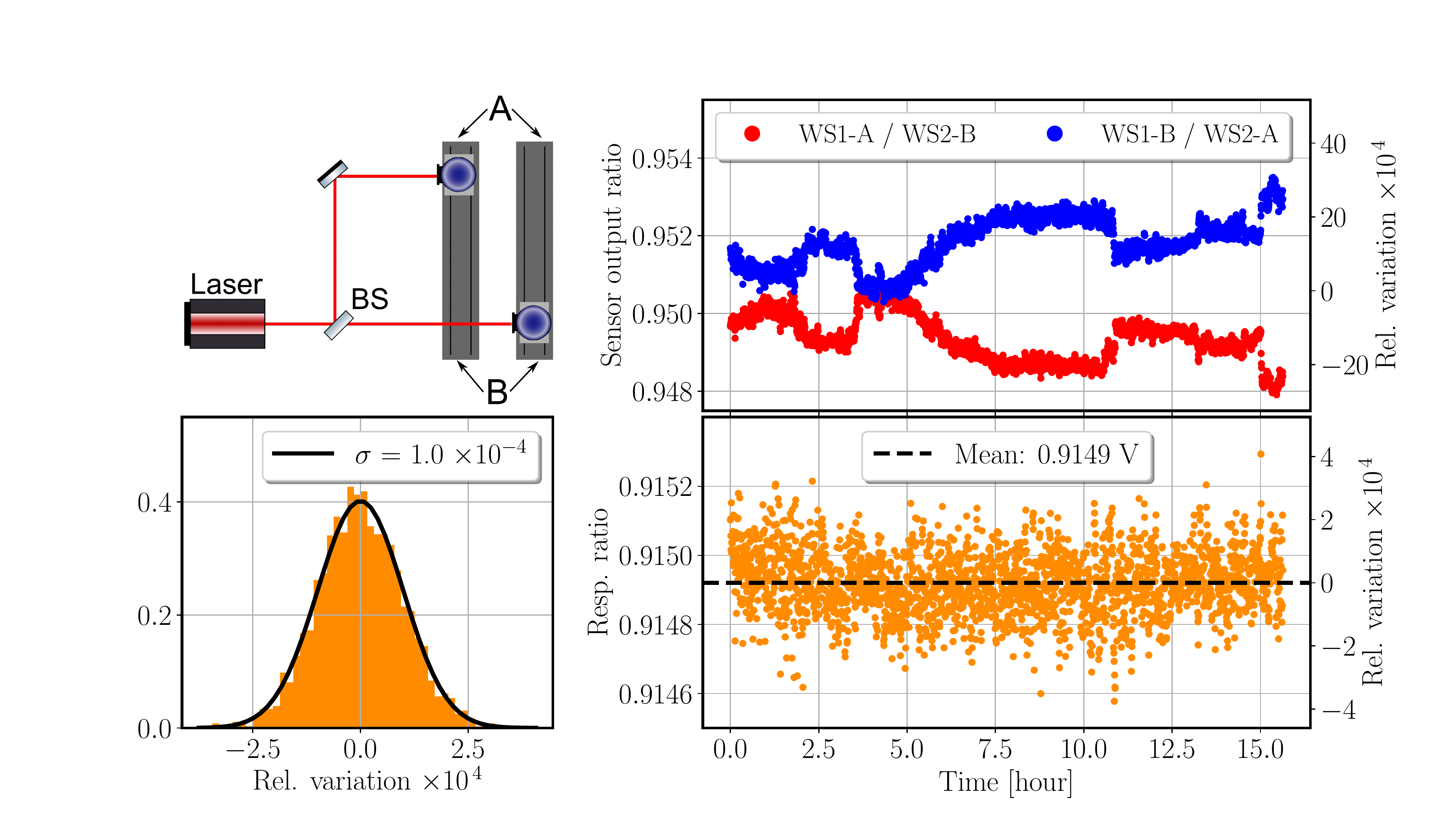}
    \caption{ Responsivity ratio measurement setup and results for two working standards. {\em Upper-left panel}: schematic diagram of laboratory setup with pneumatic slides for alternating positions of the two sensors between the beamsplitter transmitted and reflected beams; {\em Upper-right panel}: time series of ratio in A-B configuration (red) and in B-A configuration (blue); {\em Lower-right panel}: square root of the product of subsequent A-B, B-A ratios, $\alpha_{_{\textrm{W1W2}}}=\rho_{_{\mathrm{WS1}}}/\rho_{_{\textrm{WS2}}}$; {\em Lower-left panel}: normalized histogram of the relative variations in the 2800 measurements, each a from a twenty second long sequential measurement suite.}
    \label{fig:ratios_hist}
    \end{center}
\end{figure}

The GS is sent to NIST annually for calibration. %\,\cite{gs_index}.  
To transfer the GS calibration to the various working standards, a series of responsivity ratio measurements are made in a laboratory at the LHO\,\cite{in_lab_index}.  This process has been improved by using a spare Pcal transmitter module\,\cite{RSIpaper} that incorporates laser power stabilization and delivers two output beams with powers balanced to within 1\,\%.  The GS and one WS are mounted on automated pneumatic slides that alternate the positions of the two detectors between the two output beams as shown schematically in the upper-left panel of \fref{fig:ratios_hist}.
Division of the output voltages recorded simultaneously in a given configuration minimizes variations induced by laser power changes; sequential measurements with the detector positions swapped minimizes the impact of changes in the reflectivity of the beamsplitter that separates the two beams.

\begin{figure}[t]%
    \begin{center}
    \includegraphics[trim= 0.5cm 0.2cm 0.8cm 1.8cm, clip=true, width = 1\textwidth]{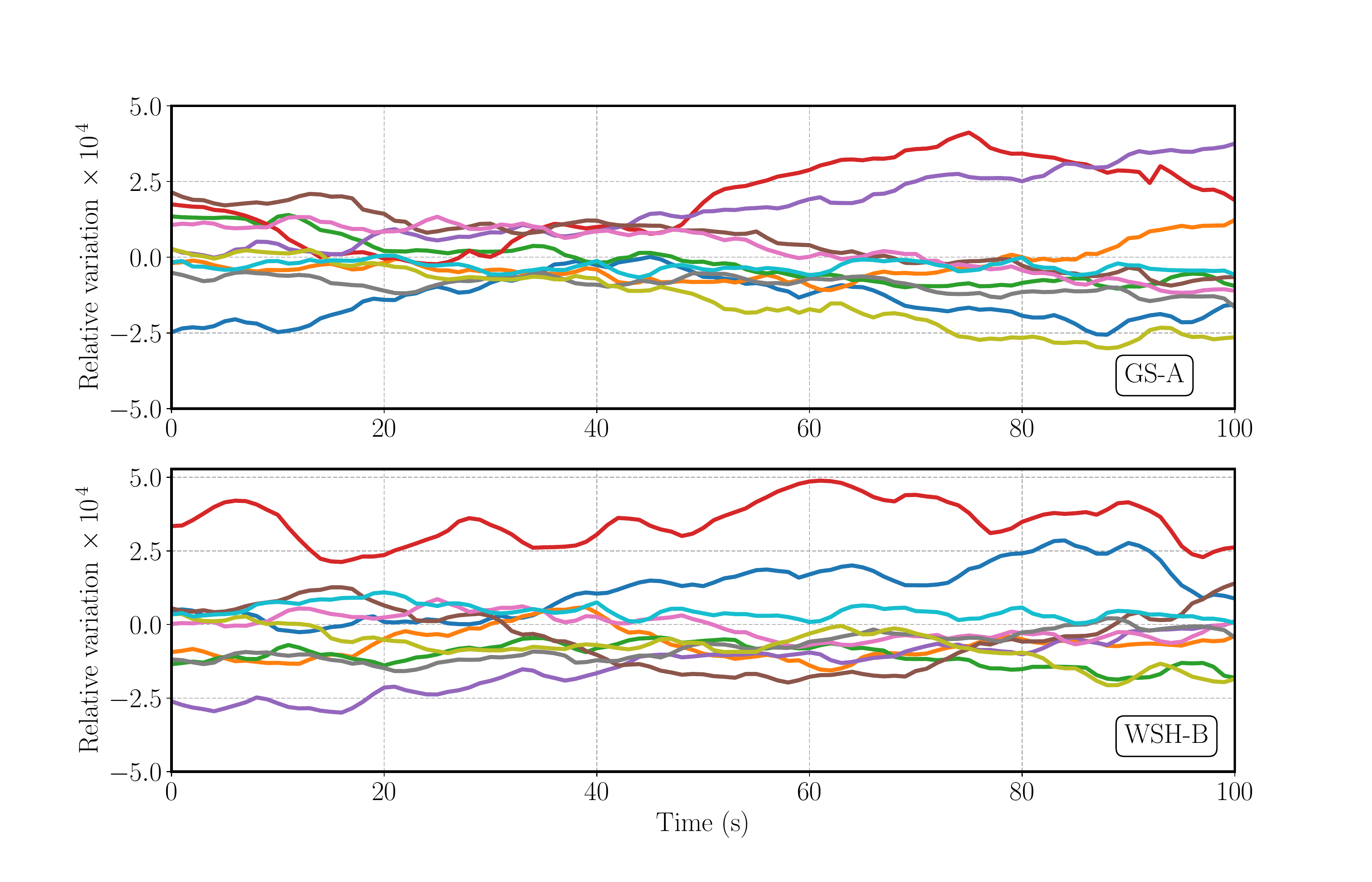}
    \caption{Laser speckle in  the outputs of the transfer standards.  {\em Upper panel}: Relative variations in ten time series of the GS transfer standard output during $\alpha_{_{WG}}$ responsivity ratio measurements.    {\em Lower panel}: Relative variations in the ten coincident WSH time series.  
    The outputs are sampled once per second and the data are normalized to the mean of all the data plotted in each panel.
    Data plotted with the same colors were recorded simultaneously.
    The GS was in position A noted in the upper-left panel of \fref{fig:ratios_hist} and WSH was in position B.}
    \label{fig:raw_voltages}
    \end{center}
\end{figure}

The transfer standards and the Rx power sensors are comprised of integrating spheres with Spectralon\textsuperscript{\textregistered} interior shells (Labsphere model 3P-LPM-040-SL) and custom-built photodetectors.
Although the integrating spheres are largely insensitive to the incident beam position, angle, polarization, and size, they exhibit laser speckle due to the coherence of the laser light that correlates the output time series \cite{speckle_text}.  This temporal correlation limits the precision of the responsivity ratio measurements and can introduce systematic errors. Figure\,\ref{fig:raw_voltages} shows ten, 100 second long time series for both the GS (upper panel) and the WSH (lower panel) standards during responsivity ratio measurements in the LHO laboratory. The temporal correlation of the data in a given time series, together with the lack of correlation with the simultaneously-recorded time series from the other sensor, is the result of laser speckle in the sensor outputs. During responsivity ratio measurements in the LHO laboratory, the impact of the laser speckle is ameliorated by recording shorter time series and swapping sensor positions more frequently.

A typical reported responsivity ratio value, $\alpha_{_{WG}}=\rho_{_{W}}/\rho_{_{G}}$, where $\rho_{_{W}}$ is the responsivity of the WS and $\rho_{_{G}}$ is that of the GS, is given by the average of 100 measurements.  The data for each measurement is comprised of four, five second long time series sampled once per second. The first two are sampled simultaneously with the detectors in the A-B configuration as shown in the upper-left panel in figure\,\ref{fig:ratios_hist}, and the last two with the detector positions swapped (B-A).  The square root of the product of the ratios of each pair of time series yields an estimate of the responsivity ratio every twenty seconds (five seconds are required to re-position the sensors).  

To elucidate this method, individual ratios (A-B and B-A configurations) for 2800 twenty second long measurement suites are plotted in the upper-right panel in figure\,\ref{fig:ratios_hist}.  Variations in the beamsplitter ratio are evident in the ``mirrored'' appearance of the two data sets.  These variations are minimized when calculating $\alpha_{_{W1W2}}$,  the square root of the product of the ratios from sequential measurements (one point for each data set), as shown in the lower-right panel of \fref{fig:ratios_hist}.  The standard deviation of the relative variation of the ratio measurements is \num{1.0e-4} as shown in the normalized histogram in the lower-left panel in figure\,\ref{fig:ratios_hist}.  Note that the Pcal power sensor responsivities have non-negligible dependence on temperature, as discussed in more detail below  and in section\,\ref{subsec:ResultSensecal}.   The data in the lower panels of figure\,\ref{fig:ratios_hist} were ``de-trended'' by correcting for the   slight variation (\SI{1.1e-4}{\per K}) in relative responsivity due to the changing laboratory ambient temperature (\SI{1.3}{K}) during this fifteen hour long measurement interval. 

To propagate the GS calibration to the Pcal power sensors at the end stations, a series of measurements are made with a working standard\,\cite{end_station_index}. They involve placing the WS alternately in the path of one or the other Pcal beam in both the transmitter and receiver modules and recording time series of the WS, Rx, and Tx power sensors using the digital data acquisition system\,\cite{P2000107}.  The end station data acquisition system is used for the WS signal, rather than the digital volt meter that is used for the responsivity ratio measurements in the laboratory setup.  This is both for convenience and to synchronize the WS data with the Rx and Tx sensor data.  Propagating the WS calibration to the end station thus involves the additional step of measuring the conversion factor, $\zeta_{_{W}}$, between volts registered by the digital volt meter and digital counts reported by the end station data system.  This is accomplished via a calibrated voltage source.
Synchronizing these time series reduces the impact of laser power variations and yields the Rx/WS responsivity ratio, $\rho_{_{R}}/\rho_{_{W}}=\alpha_{_{RW}}$.
These measurements also yield estimates of the overall optical efficiency, $\eta$.

Combining the measurements described above, the responsivity or calibration factor for the Rx end station power sensors is given by%
\begin{equation}
\label{eq:rho_TR}
    \rho_{_{R}} = \rho_{_{G}} \ \alpha_{_{WG}} \  \alpha_{_{RW}} \ \zeta_{_{W}} \,
\end{equation}
in units of ct/W.

Measurements with transfer standards are made in laboratories with differing temperatures.  Thus the temperature coefficients of the transfer standards must be taken into account when transferring laser power calibration from NIST to the sensors at the interferometer end stations. 
For a given power sensor, the temperature dependence of the responsivity can be described by 
\begin{equation}
\frac{\rho(T) } {\rho_0} =  1 + \kappa (T - T_0),
\end{equation}
where $\rho_0$ is the responsivity of the sensor measured at a reference temperature, $T_0$, and  $\kappa$ is the temperature coefficient of the relative responsivity.  By including the temperatures at which each measurement in the calibration transfer process is made, \eref{eq:rho_TR} can be rewritten as
\begin{equation}
\begin{aligned}
\label{eq:rho_TR_temp}
    \rho_{_{R}}|_{_{{T}_{_E}}} & = \rho_{_{G}}|_{_{T_{_N}}} \ \frac{\rho_{_G}|_{_{T_{_L}}}}{\rho_{_G}|_{_{T_{_N}}}} \ \frac{\rho_{_W}|_{_{T_{_L}}}}{\rho_{_G}|_{_{T_{_L}}}} \ \frac{\rho_{_W}|_{_{T_{_E}}}}{\rho_{_W}|_{_{T_{_L}}}} \  \frac{\rho_{_R}|_{_{T_{_E}}}}{\rho_{_W}|_{_{T_{_E}}}} \  \zeta_{_{W}} =
    \rho_{_{G}} \ \xi_{_{LN}} \ \alpha_{_{WG}} \ 
    \xi_{_{EL}} \ \alpha_{_{RW}} \ \zeta_{_{W}} \\ 
\end{aligned}
\end{equation}
where $T_{_{N}}$ is the NIST laboratory temperature, $T_{_{L}}$ is the  LHO laboratory temperature, and $T_{_{E}}$ is the end station temperature.  The factor $\xi_{_{LN}}$ corrects for differences in the GS responsivity measured at the NIST and LHO laboratory temperatures and the factor $\xi_{_{EL}}$ corrects for differences in the WS responsivity measured at the LHO laboratory and end station temperatures. These temperature-related correction factors can be written as
\begin{equation}
\begin{aligned}
\label{eq:xi}
    \xi_{_{LN}} = \frac{\rho_{_G}|_{_{T_{_L}}}}{\rho_{_G}|_{_{T_{_N}}}} = (1 + \kappa_{_{G}}\Delta T_{_{LN}}) \\
    \xi_{_{EL}} = \frac{\rho_{_W}|_{_{T_{_E}}}}{\rho_{_W}|_{_{T_{_L}}}} = (1 + \kappa_{_{W}}\Delta T_{_{EL}}) .
\end{aligned}
\end{equation}    
Here $\kappa_{_{G}}$ and  $\kappa_{_{W}}$ are the temperature coefficients for the GS and WS sensors normalized to their respective responsivities at their reference temperatures, $T_{_{N}}$ and $T_{_{L}}$.  $\Delta T_{_{LN}} = T_{_{L}} - T_{_{N}}$ and $\Delta T_{_{EL}} = T_{_{E}} - T_{_{L}}$\,.
The temperature-compensated Rx sensor calibration in \eref{eq:rho_TR_temp}, together with the optical efficiency factor, $\eta_{_{R}}$, given by \eref{eq:etaRetaT} and the digitized output of the Rx sensor, $d_{_{R}}$, yield the estimated power reflecting from the ETM, 
\begin{equation}
\label{eq:powerCal}
  P(\omega)  = \frac{d_{_{R}}(\omega)}{\eta_{_{R}} \, \rho_{_{R}}} =\frac{d_{_{R}}(\omega)}
  {\eta_{_{R}} \, \rho_{_{G}} \, \xi_{_{LN}} \, \alpha_{_{WG}} \, \xi_{_{EL}} \, \alpha_{_{RW}} \, \zeta_{_{W}}}\, .
\end{equation}
% %
\subsection{Calculation of displacement factors}
\label{subsec:MethodCalcs}
Combining \eref{eq:pcaldisp}, and \eref{eq:powerCal}, the digitized output of the Rx sensor can be calibrated in induced ETM displacement (units of m/ct) as
\begin{equation}
\label{eq:disp_eq}
  x(\omega)  \simeq - \frac{2 \, \rm cos\,\theta}{M \, c \, \omega^2} P(\omega) = - \frac{2 \, \rm cos\,\theta}{M \, c \, \omega^2} \frac{d_{_{R}}(\omega)}
  {\eta_{_{R}} \, \rho_{_{G}} \, \xi_{_{LN}} \, \alpha_{_{WG}} \, \xi_{_{EL}} \, \alpha_{_{RW}} \, \zeta_{_{W}}}  = - \frac{X}{\omega^2}\, d_{_{R}}(\omega)\,.
\end{equation}
Here  we define the displacement factor, $X$, as 
\begin{equation}
 X =  \frac{2 \, \rm cos\,\theta}{M \, c} \frac{1}
  {\eta_{_{R}} \, \rho_{_{G}} \, \xi_{_{LN}} \, \alpha_{_{WG}} \, \xi_{_{EL}} \, \alpha_{_{RW}} \, \zeta_{_{W}}}\, .
\label{eq:disp_eqn}
\end{equation}

Like LIGO, most GW observatories have implemented, or plan to implement, Pcals at both end stations. The Pcal systems at each end station are calibrated using the same procedures.  Laser interferometers are designed and tuned to sense differential arm length variations induced by ETM motion without regard (except for the sign of the relative displacement) to which ETM is moving.   Thus, comparing the Pcal fiducials produced at both end stations in the interferometer output signal directly measures the ratio of the Pcal calibrations at each end.  This comparison can be used to reduce the uncertainty in the induced displacements due to factors that are not common to both ends.

The X/Y calibration comparison is realized by modulating the two Pcal systems at closely-separated frequencies within the sensitive band of the interferometer.  Comparison of the amplitudes of the peaks in fast Fourier transforms (FFTs) of the interferometer output signal with the amplitudes of the associated peaks in the calibrated  Pcal end station sensor outputs yields the X/Y Pcal calibration comparison factor, $\chi_{_{XY}}$.  If there were no uncertainties in the displacement calibration factors for both end stations, $X_{_{X}}$ and $X_{_{Y}}$, this factor would be~1.  However, errors induced by uncertainties in factors that are not common to both end stations ($\rm cos\,\theta, \ M, \ \eta_{_{R}}, \ \xi_{_{EL}}, \ \alpha_{_{RW}},\ \zeta_{_{W}}$, and $\epsilon_{rot}$) cause it to deviate from~1.

Using $\chi_{_{XY}}$ and these uncertainty estimates, correction factors for each end station can be calculated.
%such that $C_{_{X}} C_{_{Y}} = \chi_{_{XY}}$.
The {\em combined} displacement factors are defined by
\begin{equation}
\label{eq:disp_coeffXY}
    X^c_{_X}  = X_{_{X}} / C_{_X}  \ \textrm{and} \  X^c_{_Y}  = X_{_{Y}}  \  C_{_Y} \,
\end{equation}
where $C_{_X}$ and $C_{_Y}$ are the X-arm and Y-arm correction factors and 
\begin{equation}
%  \frac{ X^c_{_X}} {  X^c_{_Y}} = \frac{X_{_{X}} / C_{_X}} { X_{_{Y}}  \  C_{_Y}} = 
  C_{_X} C_{_Y} \equiv  \chi_{_{XY}}.
\end{equation}
The superscript~c denotes that these displacement factors are calculated from the combination of the X-end and Y-end calibration results.  For the special case where the uncertainties are the same at both end stations, the factors are given by
$C_{_X} = C_{_Y} = \sqrt{\chi_{_{XY}}}$.

In general, the correction factors are calculated using the weighted geometric mean, $\mu_g$, of~1 and $\chi_{_{XY}}$. The weighting factors are given by the inverse of the estimated variances in the end station displacement factors, $X_{_{X}}  \ \textrm{and} \ X_{_{Y}}$, due to uncertainty contributions that are not common to both end stations.  With the $X_{_X}$ weighting factor applied to 1 and the $X_{_Y}$ weighting factor applied to $\chi_{_{XY}}$, the correction factors are given by
\begin{equation}
\label{eq:corr_fact}
    C_{_X} = \mu_g \ \textrm{and} \ C_{_Y} = \chi_{_{XY}} / \mu_g\,.
\end{equation}
The uncertainties in these factors are given by the weighted relative standard error on the geometric mean.
\section{Measurement results and uncertainty estimates}
\label{sec:results}
As an example of how the methods described in \sref{sec:method} can be applied to calibrate Pcal displacement fiducials, the results of measurements made before and during the O3 observing run are presented here.  Measurements were made for both the LHO and LLO interferometers, but for simplicity only the LHO results are presented.

The measured and calculated values of the parameters that contribute to the displacement factors described in \sref{sec:method}, and their associated relative uncertainties, are summarized in the tables below. We follow the convention used by NIST and detailed in\,\cite{NIST-1297}, employing {\em Type\,A}, {\em Type\,B}, or {\em Type\,C} evaluations for estimates of relative standard uncertainties for each parameter\footnote{ \ Evaluation of uncertainty by statistical analysis of a series of measurements is referred to as  {\em Type\,A}. Evaluation of uncertainty using means other than statistical analysis of measurements, based on scientific judgement using all relevant information available, is referred to as  {\em Type\,B.} Uncertainty estimates resulting from combinations of Type A and Type B evaluations of uncertainty are referred to as a {\em Type\,C}.}. For Type A estimates, the number of measurements is noted in parentheses after the relative uncertainties listed in the tables.
\subsection{End station power sensor calibration}
\label{subsec:ResultSensecal}
Calibration of the on-line power sensors at the end stations is realized by measurement of the six multiplicative parameters on the right-hand side of \eref{eq:rho_TR_temp}.  The GS and WS power sensors were upgraded in 2018\,\cite{ws_change}, well before the start of the O3 observing run in April 2019.  The modifications included changes in the mounting of the photodetector housing to the integrating sphere port in order to improve mechanical stability and reduce the amplitude of variations due to laser speckle.  The transimpedance amplifier was also changed to reduce complexity and minimize dark offsets.   The upgraded GS was then calibrated by NIST in December 2018.
The reported GS responsivity, $\rho_{_{G}}$, is $-8.0985$\,V/W.  
The relative uncertainty is\, 0.315\,\% (1\,$\sigma$)\,\cite{T1900097}.  This GS calibration,  reported in volts\footnote{ \ Volts reported by a dedicated GS voltmeter, Keithley model 2100.} per watt of laser radiation at 1047\,nm, is the foundation upon which the calibration of the LIGO, Virgo, and KAGRA detectors were based during the O3 observing run.

\begin{figure}[t]%
    \begin{center}
    \includegraphics[trim= 1.8cm 0.3cm 1.2cm 1.8cm, clip, width = 1.0\textwidth]{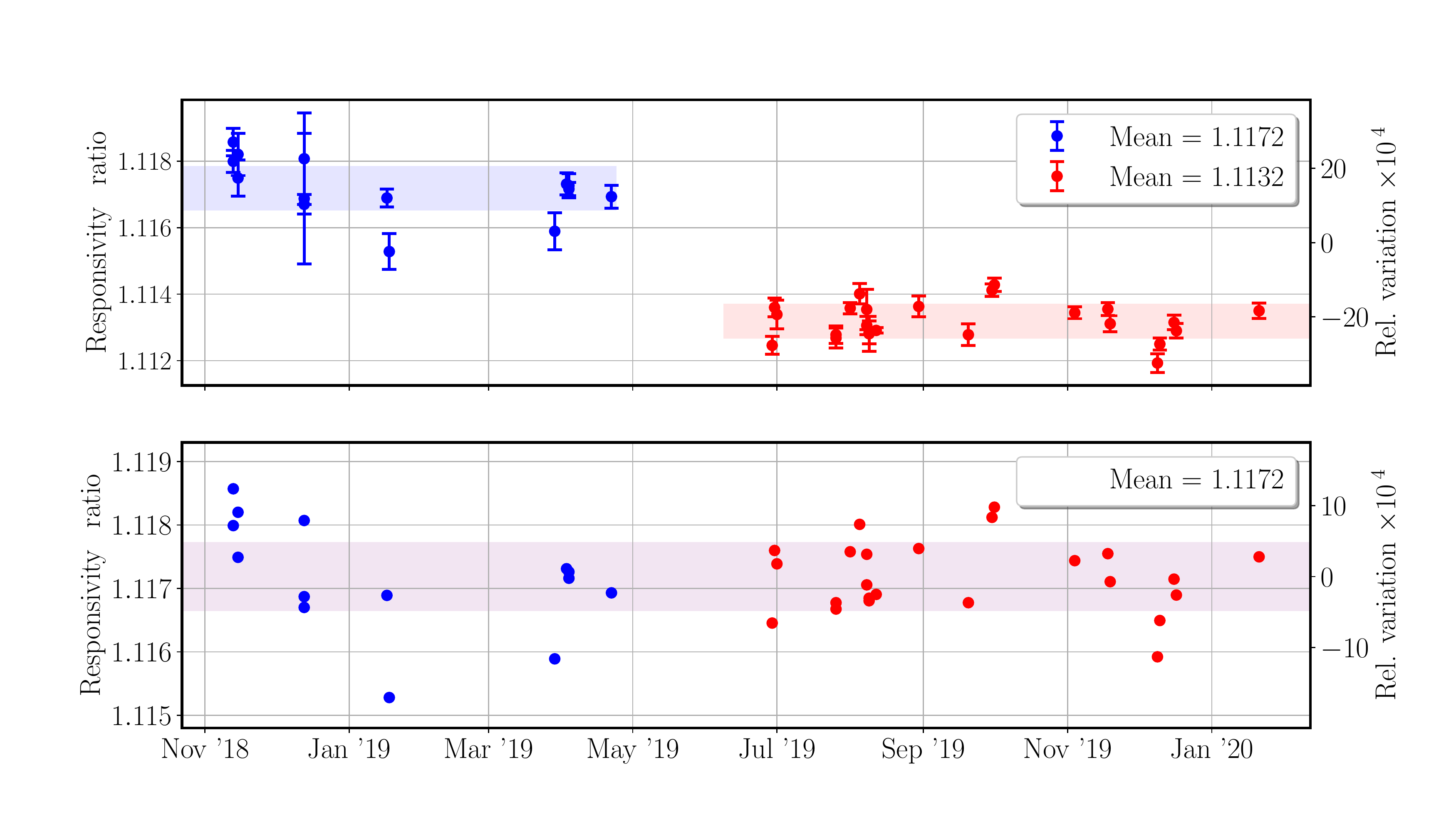}
    \caption{WSH to GS responsivity ratio, $\alpha_{_{WG}}$, measured between November 2018 and February 2020. {\em Upper panel}: The data plotted in red were measured after June 2019 when the Pcal responsivity measurement setup was moved to another building.  The colored bands indicate $\pm$ 1\,standard deviation for each measurement group. The error bars, $\pm$\,1\, standard error on the mean for each  data point have been magnified by a factor of twenty to increase visibility.  Note that they are not used for weighting, as explained in the text. {\em Lower panel}: Same data as upper panel, but with the red data points shifted by the ratio of the means of the blue and red data sets (multiplied by 1.00359).  The step in the data is attributed to a change in the GS responsivity that occurred during the move.}
    \label{fig:alphaHG}
    \end{center}
\end{figure}

As shown schematically in figure\,\ref{fig:calib_transfer}, responsivity ratio measurements performed at the LHO\,\cite{in_lab_index} are used to transfer the GS calibration to the array of working standards, WSH, WSL, WSK, WSV (and eventually WSI).  Also, responsivity ratio measurements between one or more of the working standards and the GS, both before shipping the GS to NIST and immediately after its return from calibration at NIST, are used to determine whether changes occurred during shipping.
\Fref{fig:alphaHG} shows the WS to GS responsivity ratio, $\alpha_{_{WG}}$, for WSH over the period from November 2018 to February 2020.
The mean of the four measurements made immediately before sending the GS to NIST for calibration in mid-November 2018 differs from the mean of the three measurements made immediately after it was returned  in mid-December 2018 by 0.06\,\%.  This indicates that changes during shipping were minimal.

\begin{figure}[t]%
    \begin{center}
\includegraphics[trim= 0cm 0.5cm 0cm 0cm, clip,width = 1.0\textwidth]{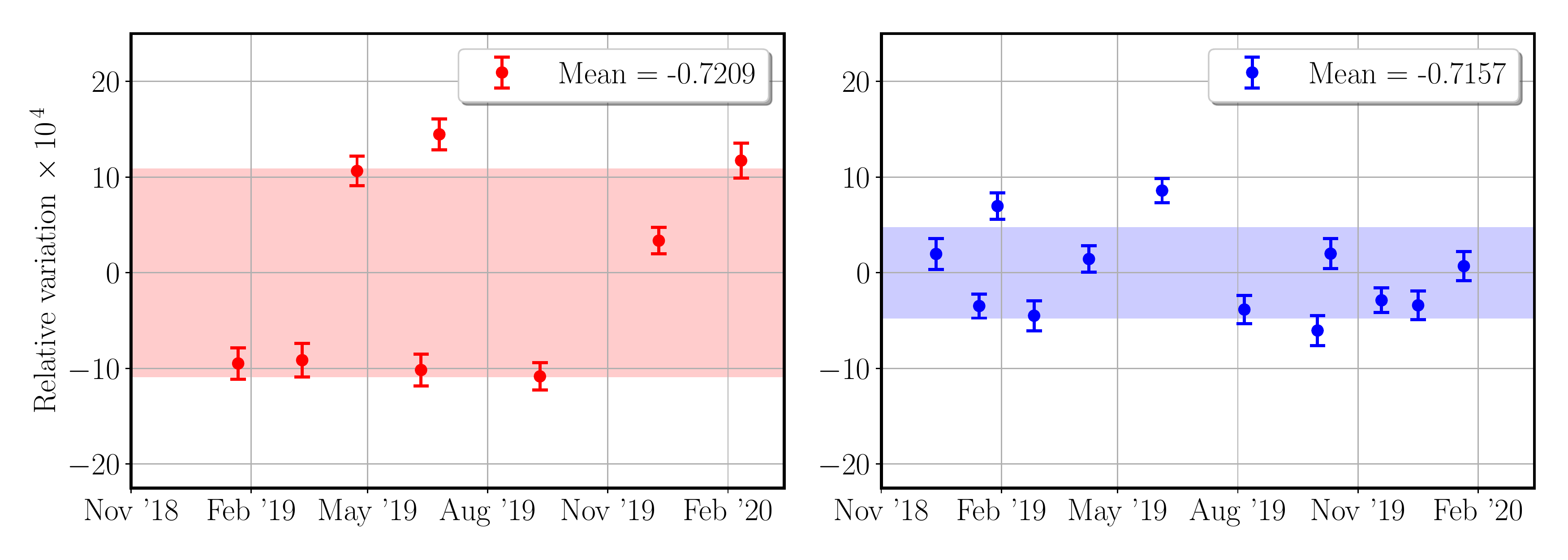}
    \caption{Relative variation of the responsivity ratio, $\alpha_{_{RW}}$, of the Rx and WSH power sensors, measured at the LHO X-end ({\em left panel}) and Y-end ({\em right panel}) stations. The shaded regions are $\pm$\,1\,standard deviation about the weighted mean values listed in the legends.  The errors bars are estimated using the formalism detailed in Appendix~A of~\cite{P1900127}.}
    \label{fig:alphaRWLHO}
    \end{center}
\end{figure}

However, as seen in the upper panel of \fref{fig:alphaHG}, a significant change in $\alpha_{_{WG}}$ (0.36\,\%) occurred in June 2019 when the responsivity ratio measurement setup was moved to a different building at the LHO.  Measurements of the responsivity ratio between WSH and the Rx sensors at the LHO end stations (see \fref{fig:alphaRWLHO}) indicate that the change in $\alpha_{_{WG}}$ was caused by a change in the GS responsivity, $\rho_{_{G}}$, alone.  The exact cause of this increase in responsivity has not yet been identified.  The lower panel in \fref{fig:alphaHG} shows the same data as in the upper panel, but with the data plotted with the red points shifted up (multiplied by 1.00359). 

The error bars in the upper panel have been magnified by a factor of twenty for better visibility. But, they have not been used for weighting because the variations in the data, each point the mean of a set of between 25 and 1200 measured values, are dominated by systematic variations, not the statistical variations of the measurements within each set.  Weighting by the inverse of the square of the standard errors on the means would thus bias the estimate of the overall mean in favor of suites with larger numbers of measurements.   The mean and the relative uncertainty of $\alpha_{_{WG}}$ for WSH, from the lower panel of \fref{fig:alphaHG}, in which the  error bars have been omitted,  are listed in \tref{tab:rho_values}. 

\begin{table}[t]
\caption{Measured responsivities of the Pcal end station power sensors, $\rho_{_{R}}$, together with contributing factors (indented) and uncertainties, for the LHO interferometer during the O3 observing run.  For Type A uncertainties, the  number of measurements is noted in parentheses.}
\vspace{0.1in}
\begin{indented}
\item[]\begin{tabular}{L{1.3cm} C{1.5cm} C{1.6cm} C{1.5cm} C{1.7cm} C{1cm} C{1cm} }
    \Xhline{4\arrayrulewidth}
     \multirow{2}{*}{\bf Param} & \multicolumn{2}{c}{LHO X-end} & \multicolumn{2}{c}{LHO Y-end} & \multirow{2}{*}{\bf Units} & \multirow{2}{*}{\bf Type}\\
    \cline{2-3} \cline{4-5}
     &\bf  Values & $\bf u_{rel}$ \bf (\%) & \bf Values & $\bf u_{rel}$ \bf (\%) & & \\
    \Xhline{2\arrayrulewidth}
     $\rho_{_R}$ & \num{1.068e4}  & 0.328 & \num{1.061e4} &  0.326  & $\rm ct/W$ & C \\
    % \hline  
    \hspace{5 mm}$\rho_{_G}$ & -8.0985 & 0.315 & \multicolumn{2}{c}{Common with X-end} & V/W & C\\
    % \hline
    \hspace{5 mm}$\alpha_{_{WG}}$  & 1.1172 & 0.010 (38) & \multicolumn{2}{c}{Common with X-end} &- & A  \\
    % \hline
    \hspace{5 mm}$\alpha_{_{RW}}$   & -0.7209  & 0.042 (8)  & -0.7157 &  0.014 (12) & - & A\\
    % \hline
    \hspace{5 mm}$\zeta_{_{W}}$   & 1636.9  &  0.002 (8)& 1637.6 & 0.002 (9) & $\rm ct/V$ & A \\
    % \hline
    \hspace{5 mm}$\xi_{_{LN}}$  & 1.0020  & 0.070  & \multicolumn{2}{c}{Common with X-end} & - & C\\
    % \hline
    \hspace{5 mm}$\xi_{_{EL}}$  & 0.9986 & 0.028 & 0.9986  &0.028 & - & C\\
    \Xhline{4\arrayrulewidth}
\end{tabular}
\end{indented}
\label{tab:rho_values}
\end{table}
To calibrate the on-line Pcal Rx and Tx power sensors at the end stations, a series of responsivity ratio measurements using a WS are performed\,\cite{end_station_index}. The Rx to WSH responsivity ratios, $\alpha_{_{RW}}$, for the LHO X-end and Y-end sensors measured between  December 2018 and March 2020, are shown in  \fref{fig:alphaRWLHO}.
The error bars are estimated from the 240 second long time series recorded for each element of the measurement suite, using the formalism described in Appendix~A of\,\cite{P1900127}.  This method uses the standard deviations of the data sets rather than estimating standard errors on the mean values,  because the data is correlated due to laser speckle.  The measured values of $\alpha_{_{RW}}$ and their  relative uncertainties, for both end stations,  are listed in \tref{tab:rho_values}.

For measurements of $\alpha_{WG}$, a digital voltmeter (Keithley Model 2100) is used to record the output of the WS. An analog-to-digital (ADC) converter that is part of the LIGO data acquisition system is used for measurements of $\alpha_{_{RW}}$ at the end stations.  The factor that converts the volts measured in the laboratory to the digital counts measured at the end station, $\zeta_{_{W}}$ in \eref{eq:rho_TR}, is measured using a calibrated voltage source (Martel Model IVC-222HPII). Before going to the end station, the ratio of the Martel output to that of the Keithley voltmeter is measured.  The Martel voltage source is then used to measure the volt-to-count conversion factor for the ADC of the WS data acquisition system  channel.  The combination of these two measurements yields $\zeta_{_{W}}$ in units of ct/V. These factors are close to, but vary slightly from, the ideal value of 1638.4 ct/V.  The measured values of $\zeta_{_{W}}$ and their relative uncertainties are listed in \tref{tab:rho_values}.

The remaining two factors on the right-hand side of \eref{eq:rho_TR_temp} are $\xi_{_{LN}}$ and $\xi_{_{EL}}$.  These correction factors account for the temperature dependence of the responsivities of the GS and WS power sensors and the  differences between the NIST laboratory, the LHO laboratory,  and the end station ambient temperatures.  To measure the temperature dependence of the WS responsivity, a temperature sensor (Analog Devices, AD590) was bonded to the photodetector transimpedance amplifier circuit board.  The WS was then placed in an oven and heated to about 7~K above the ambient laboratory temperature.  It was then moved to the responsivity ratio measurement setup and measurements of the WS/GS responsivity ratio were made as the WS cooled to room temperature.

\begin{figure}[t]
    \begin{center}
    \includegraphics[trim= 0.4cm 0.4cm 0.4cm 0.4cm, clip, width =1.0\textwidth]{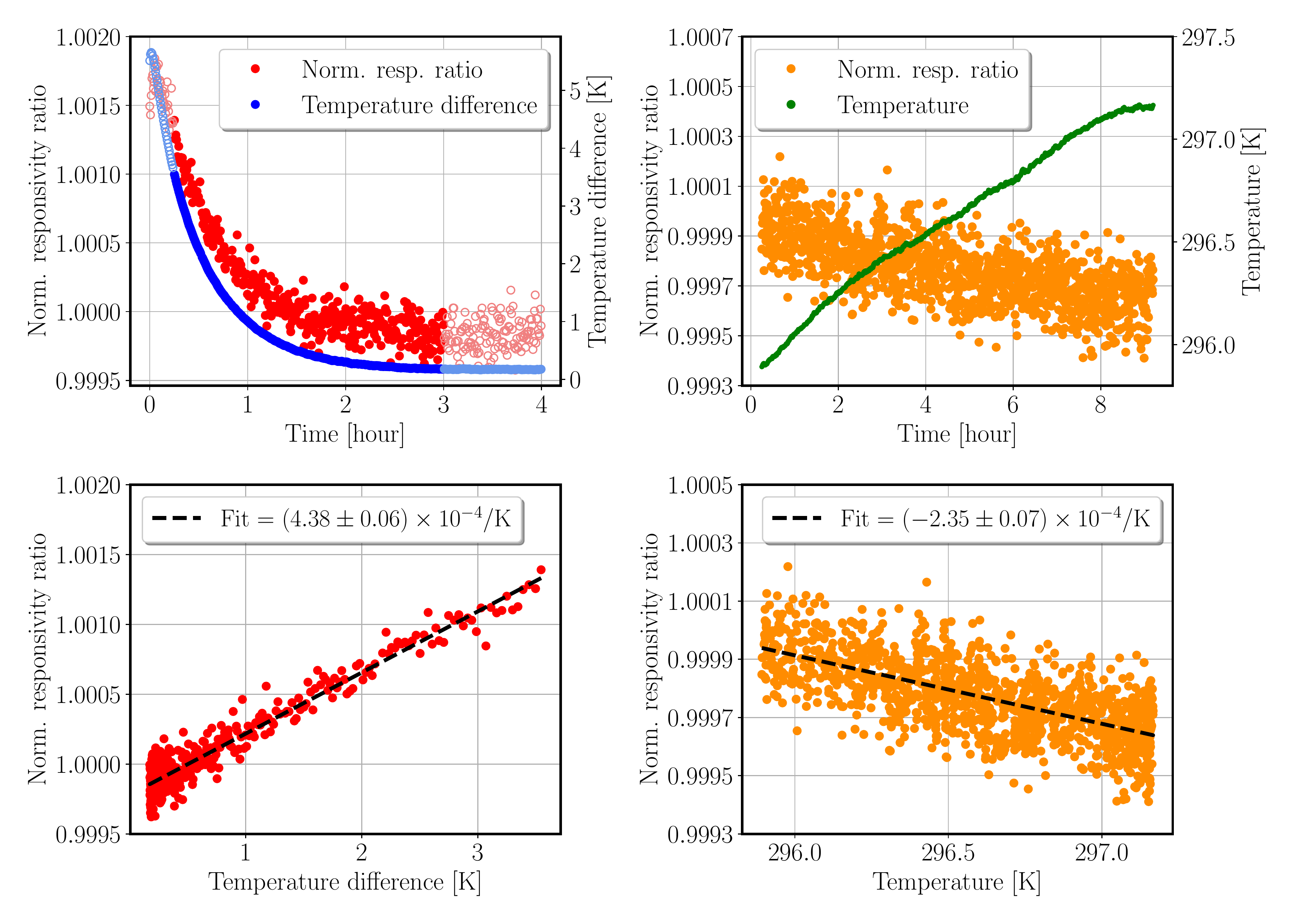}
    \caption{Temperature-dependent responsivity ratio measurements enabling estimation of the WS and GS temperature coefficients, $\kappa_{_W}$ and $\kappa_{_G}$. In all panels the responsivity ratio, $\alpha_{_{WG}}$, is normalized to its mean value of 1.1172 (see \fref{fig:alphaHG}). {\em Upper-left panel}: Normalized responsivity ratio and temperature difference between the WS and GS versus time as the WS cools after being heated in an oven.  {\em Lower-left panel}: Linear least-squares fit to the data in the upper-left panel.  The slope and uncertainty are listed in the legend.  {\em Upper-right panel}: Normalized responsivity ratio and the ambient laboratory temperature plotted versus time when both sensors were at the ambient temperature that varied more than 1\,K over almost ten hours.  {\em Lower-right panel}: Linear least-squares fit to the data in the upper-right panel with slope and uncertainty listed in the legend.}
    \label{fig:tempV1}
    \end{center}
\end{figure}

The upper-left panel of \fref{fig:tempV1} shows  $\alpha_{_{WG}}$, normalized to the mean value of 1.1172 (see \fref{fig:alphaHG}), and the difference between the WS and GS temperatures plotted versus time as the WS cools.  The lower-left panel shows a linear, least-squares fit to the data indicating a relative responsivity temperature coefficient for WS, $\kappa_{_{W}}$, of \num{4.38e-4}/K. The uncertainty in the fit is \num{0.06e-4}/K. 

To investigate the temperature coefficient of the GS, we recorded data with both the WS and the GS at the laboratory ambient temperature during a period when the ambient temperature varied by about 1.3\,K.  The laboratory ambient temperature and $\alpha_{_{WG}}$ normalized to 1.1172  are plotted in the upper-right panel in \fref{fig:tempV1}.  A  linear, least-squares fit to the data, yielding a slope of \num{-2.35e-4}/K with uncertainty of \num{0.07e-4}/K,  is shown in the lower-right panel.  Using $\kappa_{_{W}}$, from the data in the left panels of \fref{fig:tempV1},  the inferred coefficient for the GS, $\kappa_{_{G}}$, is \num{6.73e-4}/K. 

Temperature differences between the NIST, the LHO, and end station measurement environments were quantified using a set of digital thermometers that were huddled to assess relative offsets, then deployed to each location.  The NIST measurement laboratory temperature was set to \SI{20}{\celsius}, the mean temperature of the LHO responsivity ratio measurement laboratory was \SI{23}{\celsius}, and the mean X-end and Y-end temperatures were \SI{19.8}{\celsius} and \SI{19.7}{\celsius} respectively. The end station temperatures varied by about $\pm$\,\SI{0.5}{\celsius} over six months. The calculated values for $\Delta T_{_{LN}}$ and $\Delta T_{_{EL}}$, together with the measured GS and WS temperature coefficients, $\kappa_{_G}$ and $\kappa_{_W}$, are used to calculate $\xi_{_{LN}}$ and $\xi_{_{EL}}$ using \eref{eq:xi}.  Their values and associated relative uncertainties are listed in \tref{tab:xi_values}.
\begin{table}[t]
\caption{Measured temperature correction factors, $\xi_{_{LN}}$ and $\xi_{_{EL}}$, for the Pcal end station power sensor calibrations,  together with contributing factors (indented) and uncertainties, for the LHO interferometer during the O3 observing run.}
\vspace{0.1in}
\begin{indented}
\item[]\begin{tabular}{L{1.5cm} C{1.5cm} C{1.5cm} C{1.5cm} C{1.5cm} C{1cm} C{1cm} }
        \Xhline{4\arrayrulewidth}
        \multirow{2}{*}{\bf Param} & \multicolumn{2}{c}{LHO X-end} & \multicolumn{2}{c}{LHO Y-end} & \multirow{2}{*}{\bf Units} & \multirow{2}{*}{\bf Type}\\
        \cline{2-3} \cline{4-5}
         &\bf  Values & $\bf u_{rel}$ \bf (\%) & \bf Values & $\bf u_{rel}$ \bf (\%) & & \\
        \Xhline{2\arrayrulewidth}
         $\xi_{_{LN}}$ & 1.0020  & 0.070 & \multicolumn{2}{c}{Common with X-end} &- & C\\
        
        \hspace{5 mm}$\kappa_{_G}$  & \num{6.73e-4} & 1.4 & \multicolumn{2}{c}{Common with X-end} & 1/K & A \\
        
        \hspace{5 mm}$\Delta T_{_{LN}}$  & 3.0 & 34 & \multicolumn{2}{c}{Common with X-end} & K & C \\
        
        $\xi_{_{EL}}$ & 0.9986 & 0.040 & 0.9986 & 0.040 & -& C\\
        
        \hspace{5 mm}$\kappa_{_W}$ &  \num{4.38e-4} & 1.4 &  \multicolumn{2}{c}{Common with X-end} & 1/K & A \\
        
        \hspace{5 mm}$\Delta T_{_{EL}}$  & -3.2 & 28 & -3.3 & 28 & K & C \\ 
        \Xhline{4\arrayrulewidth}
        \end{tabular}
\label{tab:xi_values}
\end{indented}
\end{table}
 \subsection{Displacement factors}\label{subsec:ResultsForcecoeff}
\begin{figure}[t]%
    \begin{center}
\includegraphics[trim= 0cm 0.5cm 0cm 0cm, clip,width = 1.0\textwidth]{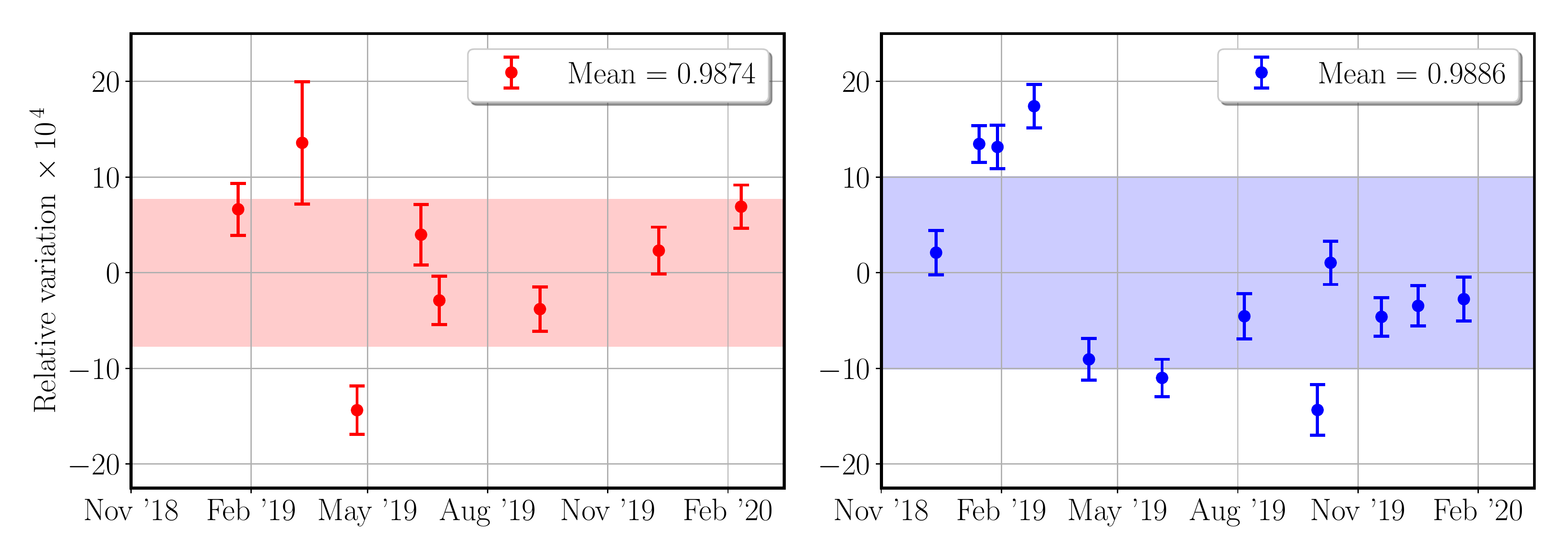}
    \caption{Relative variation of the optical efficiency between the transmitter and receiver modules at the end stations, $\eta$, measured at the LHO X-end ({\em left panel}) and Y-end ({\em right panel}) stations. The shaded regions are $\pm$\,1\,standard deviation about the weighted mean values listed in the legends. The errors bars are estimated using the formalism detailed in Appendix~B of~\cite{P1900127}.}
    \label{fig:etaLHO}
    \end{center}
\end{figure}

As noted in \sref{subsec:MethodSensorcal}, the end station measurements made with the WS also yield measurements of the overall optical efficiency, $\eta$,  for propagation of the laser beams between the Tx and Rx modules.  Assuming that $\beta$ is constant, using \eref{eq:etaRetaT} the WS end station measurements yield estimates of the optical efficiency, $\eta_{_{R}}$, required to calibrate the end station power sensors in terms of laser power reflecting from the ETM using \eref{eq:power_prime}.   The measured values of $\eta$ for the LHO X-end and Y-end stations from November 2018 to February 2020 are plotted in \fref{fig:etaLHO}.  The error bars, used for calculating weighted values, are generated using the formalism detailed in Appendix~B of~\cite{P1900127}.
The mean values of $\eta_{_{R}}$, $\beta$, and $\eta$, together with their relative uncertainties, are listed in \tref{tab:eta_values}.
\begin{table}[t]
\caption{Measured optical efficiency correction factors, $\eta_{_{R}}$,  for the receiver-side end station power sensors,  together with contributing factors (indented) and uncertainties, for the LHO interferometer during the O3 observing run. For Type\,A uncertainties, the  number of measurements is noted in parentheses.}
\vspace{0.1in}
\begin{indented}
\item[]\begin{tabular}{L{1.5cm} C{1.5cm} C{1.5cm} C{1.5cm} C{1.5cm} C{1cm}}
        \Xhline{4\arrayrulewidth}
         \multirow{2}{*}{\bf Param} & \multicolumn{2}{c}{LHO X-end} & \multicolumn{2}{c}{LHO Y-end} & \multirow{2}{*}{\bf Type}\\ \cline{2-3} \cline{4-5}
         &\bf  Values & $\bf u_{rel}$ \bf (\%) & \bf Values & $\bf u_{rel}$ \bf (\%)  & \\
        \Xhline{2\arrayrulewidth}
         $\eta_{_{R}}$ & 0.9942 & 0.04 & 0.9948  & 0.04 &  C \\ 
        
        \hspace{5 mm}$\eta$ & 0.9874 & 0.03 (8) & 0.9886  & 0.03 (12) &  A\\
        
        \hspace{5 mm}$\beta$ & 0.9989 &  0.08 (3) & 0.9988  &  0.08 (3) &  A\\
        \Xhline{4\arrayrulewidth}
    \end{tabular}
\label{tab:eta_values}
\end{indented}
\end{table}

Using right hand side of \eref{eq:pcaldisp}, in which suspension displacement and rotation transfer functions have been approximated by the responses of a free mass, only the mass of the ETM is required to convert the force given by \eref{eq:force} to the displacement factor, $X$, in \eref{eq:disp_eq}.  Here, we are neglecting the second term in square brackets because its estimated magnitude is treated as an uncertainty, $\epsilon_{rot}$.  The ETM masses are measured by the vendor that polishes the mirrors and at the end stations before they are suspended.  The calibration of the electronic balances used before suspending have been verified using two \SI{20}{\kilogram} calibrated reference masses.  The maximum estimated uncertainty in the $\sim$\,\SI{40}{\kilogram} suspended mass is \SI{10}{\gram}.

The angle of incidence, $\theta$, is determined by the last relay mirrors on the periscope structure located inside the vacuum envelope\,\cite{RSIpaper}.  These mirrors direct the Pcal beams such that they impinge on the ETM at positions above and  below center, on the vertical center line of the face of the ETM. The nominal angle of incidence is 8.72~degrees. The maximum deviations of this angle are bounded by the size of the periscope optics (2~inch diameter) that relay the beams to the end test mass.

When the interferometer beam is not centered on the face of the ETM, as is currently the practice at the LIGO observatories to mitigate the impact of point absorbers in the mirror coatings\,\cite{pointDefects}, Pcal-induced  rotations can increase or decrease the sensed ETM displacement.  Force-to-angle coupling investigations using the ETM orientation actuators have been used to infer the interferometer beam offsets, but for Pcal beam position offsets we currently only have estimates of their maximum magnitudes.

\begin{figure}[t]%
    \begin{center}
    \includegraphics[width = 0.9\textwidth]{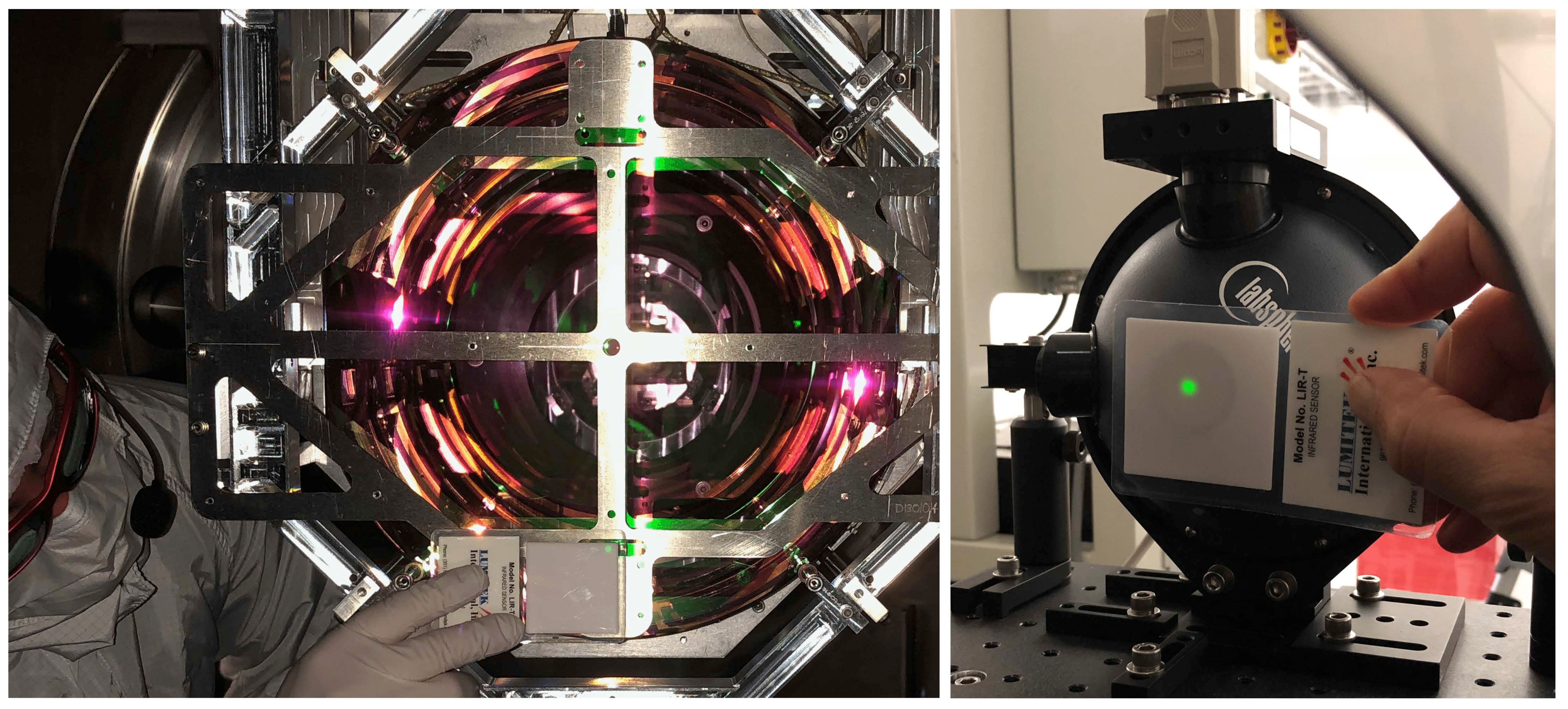}
    \caption{{\em Left image}: Alignment of Pcal beams using a target mounted to the suspension frame of the ETM.  {\em Right image}: Checking Pcal beam spots positions at the entrance aperture of the Rx power sensor.}
    \label{fig:PcalBeams}
    \end{center}
\end{figure}

The Pcal beams are positioned on the ETM surface when the vacuum envelope is vented, using targets that are bolted to the suspension structure surrounding the ETM (see left image in \fref{fig:PcalBeams}).  The reflected beams exiting the vacuum envelope are aligned to the center of the entrance aperture of the Rx power sensor (see right image in \fref{fig:PcalBeams}).
Monitoring of the locations of the beams within the aperture, $\sim$\,\SI{12}{\meter} from the Tx module output, indicates that the maximum displacement of the beams from their optimal locations on the ETM surface is $\pm$\,2~mm.  The powers in the two beams are balanced to within \SI{1}{\percent}. With their nominal locations 111.6~mm from the center of the ETM surface,  this power imbalance causes the center of force to be displaced by an additional $\sim$\,0.5~mm. The magnitude of $\vec{a}$ is estimated by adding the Pcal beam position offset and power imbalance contributions in quadrature.

The interferometer beam position offsets from center are 29\,mm for the X-end ETM and 22\,mm for the Y-end ETM.  The uncertainty introduced by unintended rotation of the ETM, $\epsilon_{rot}$, is proportional to the dot product of the Pcal and interferometer  beam offset vectors, $\vec{a}$ and $\vec{b}$, i.e. $\epsilon_{rot} \propto |\vec{a}| |\vec{b}| \cos\phi$.     Because $\phi$, the angle between $\vec{a}$ and $\vec{b}$, is equally probable to be any value between -$\pi$ and $\pi$, we use a {\em sine wave}, or {\em U-shaped}, probability density function\,\cite{bendat} to estimate the variance in $ \cos\phi$ and form a Type\,B estimate for $\epsilon_{rot}$ using \eref{eq:pcaldisp},  $\epsilon_{rot} = Mab/(\sqrt{2}I)$.   The values of the relative uncertainty estimates for both the X-end and Y-end stations at LHO are listed in \tref{tab:rot_values}. Because of these large interferometer beam position offsets,   $\epsilon_{rot}$ is currently one of the largest sources of uncertainty for the LIGO Pcal systems.
\begin{table}[t]
\caption{Estimated uncertainties due to unintended rotation of the ETM induced by Pcal forces, $\epsilon_{rot}$, together with contributing factors (indented),  for the LHO interferometer during the O3 observing run.}
\vspace{0.1in}
\begin{indented}
    \item[]\begin{tabular}{L{1.5cm} C{2.5cm} C{2.5cm} C{1.5cm}}
        \Xhline{4\arrayrulewidth}
         \bf Param & {\bf LHO-X} & {\bf LHO-Y} & {\bf Units}  \\
        \Xhline{2\arrayrulewidth}
        $\epsilon_{rot}$ & 0.41\,\% & 0.31\,\% & -  \\
        
         \hspace{5 mm}$|\vec{a}|$  & \num{2e-3} & \num{2e-3}  & m \\
        
        \hspace{5 mm}$|\vec{b}|$  & \num{22e-3} & \num{29e-3} & m \\
        
        \hspace{5 mm}$I_p$  & 0.419 & 0.419 & $\rm kg\,m^2$ \\
        
        \hspace{5 mm}$I_y$  & 0.410 & 0.410 & $\rm kg\,m^2$\\
        
        \hspace{5 mm}$M/I_p $   & 94.65 & 94.47  & $\rm 1/m^{2}$\\
        
        \hspace{5 mm}$M/I_y$   & 96.68 &96.50  & $\rm 1/m^{2}$ \\
        \Xhline{4\arrayrulewidth}
        \end{tabular}
\label{tab:rot_values}
\end{indented}
\end{table}

Combining the factors discussed above, the displacement factors for each end station are calculated using \eref{eq:disp_eqn}.   The values of  $X$ for both end stations, together with their relative uncertainties are listed in \tref{tab:Xfactor}.  The overall uncertainties are dominated by the uncertainties due to the unintended rotation of the ETM, $\epsilon_{rot}$, and the uncertainty in the calibration of the end station power sensors, $\rho_{_{R}}$.  
The relative uncertainties of 0.53\,\% for the X-end and 0.45\,\% for the Y-end are smaller than the lowest values previously reported, 0.75\,\%\,\cite{RSIpaper}. 
\begin{table}[t]
\caption{Measured Pcal displacement factors, together with contributing factors (indented) and uncertainties, for the LHO interferometer during the O3 observing run.}
\vspace{0.1in}
\begin{indented}
    \item[]\begin{tabular}{L{1.5cm} C{2cm} C{1.2cm} C{2cm} C{1.2cm} C{1cm} C{0.7cm} }
        \Xhline{4\arrayrulewidth}
        \multirow{2}{*}{\bf Param}& \multicolumn{2}{c}{ LHO X-end} & \multicolumn{2}{c}{ LHO Y-end} & \multirow{2}{*}{\bf Units} &\multirow{2}{*}{\bf Type} \\
        \cline{2-3} \cline{4-5}
        & {\bf Values} &  ${\bf u_{rel}}$\,{\bf(\%)} & {\bf Values} & ${\bf u_{rel}}$\,{\bf(\%)} & & \\
        \Xhline{2\arrayrulewidth}
        $X_{_X}, X_{_Y}$  & \num{1.565e-14} & 0.53 &  \num{1.578e-14} & 0.45  &  m/ct & C\\
       \hspace{5mm}$\cos \theta$   & 0.9884 & 0.03  & 0.9884 & 0.03 &- & B\\
        \hspace{5mm}$M$ & 39.657 & 0.01 & 39.584  & 0.01  & kg & B  \\
      
        \hspace{5mm}$\epsilon_{rot}$  & - & 0.41 &- & 0.31  &- & B \\
  
        \hspace{5mm}$\rho_{_R}$  & \num{1.068e4}  & 0.33 & \num{1.061e4} & 0.33 & ct/W & C \\
     
        \hspace{5mm}$\eta_{_{R}}$  & 0.9942 & 0.04 & 0.9948 & 0.04 &- & C \\ 
        \Xhline{4\arrayrulewidth}
        \end{tabular}
\label{tab:Xfactor}
\end{indented}
\end{table}

\begin{figure}[t]%
    \begin{center}
    \includegraphics[width = 1.0\textwidth]{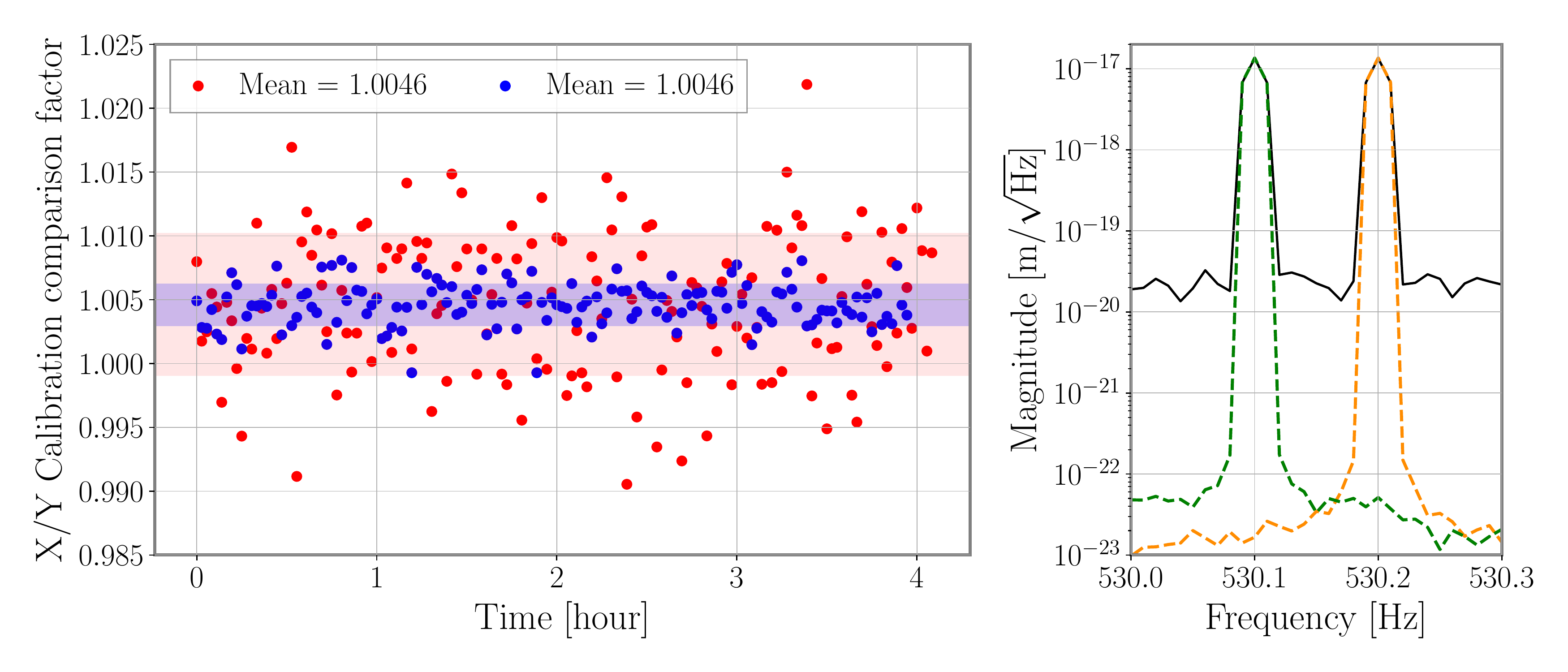}
    \caption{{\em Left panel}: $\chi_{_{XY}}$, the ratio of the amplitudes of the displacements reported by the calibrated X-end and Y-end Pcal Rx sensor signals, divided by the ratios of the amplitudes of the peaks in the interferometer output signal. The red points, for data recorded on February 3, 2020, are with the X-end excitation at 530.1\,Hz and the Y-end  at 530.2\,Hz; the blue points, for data recorded on March 2, 2020, are with the X-end and Y-end excitation frequencies swapped and with the higher excitation amplitudes. The shaded regions are $\pm$\,1\,standard deviation about the mean values. {\em Right panel}: Amplitude spectral density of the calibrated Pcal X-end (orange) and Y-end (green) Rx sensor outputs and of the interferometer output signal (black).  The measurement bandwidth is 0.01\,Hz. }
    \label{fig:X_vs_Y}
    \end{center}
\end{figure}

To compare the Pcal calibrations at the two end stations, large-amplitude displacements at frequencies separated by 0.1~Hz were induced by the X- and Y-arm Pcals as shown in the right panel in \fref{fig:X_vs_Y}.
The calibrated Rx power sensors at each end station provide estimates of the induced displacement amplitudes (see \eref{eq:disp_eq}).  The induced periodic displacements appear with high SNR in the interferometer output signal and the Rx power sensor signals as shown in the right panel of \fref{fig:X_vs_Y}.  The Spectral Line Monitoring tool\,\cite{SLMTool} was used to perform long-duration (100~s) FFT measurements of the line amplitudes in the interferometer and Rx sensor output time series.   
The left panel of \fref{fig:X_vs_Y} shows $\chi_{_{XY}}$, the ratio (X/Y) of the  amplitudes of the calibrated X-end and Y-end Pcal Rx sensor output signals, each normalized to the amplitude of the respective peak in the interferometer output signal. 
To ensure that small variations in the interferometer response function over the 0.1~Hz frequency separation between the two excitations were not impacting the comparison, the excitation and analysis were repeated with the frequencies of the X-end and Y-end excitations swapped.  Increasing the excitation amplitudes for this second comparison reduced the variations in the FFT results.  The measured value of $\chi_{_{XY}}$ for the higher SNR data set (blue points in the left panel of \fref{fig:X_vs_Y}) is 1.00460 with a relative standard uncertainty of 0.014\,\% and for the other data set (red points in the left panel of \fref{fig:X_vs_Y}) the mean value is 1.00463 with a relative standard uncertainty of 0.046\,\%.
 
The contributions to the relative uncertainties in $X$ from factors that are not common to both end stations (see \sref{subsec:MethodCalcs}) are \SI{0.42}{\percent} for X-end and \SI{0.32}{\percent} for Y-end.  Summing these uncertainties in quadrature, the calculated relative uncertainty for the quotient of the displacement factors, $X_{_X} / X_{_Y}$, is \SI{0.52}{\percent}.  Thus the measured value of 1.0046 for $\chi_{_{XY}}$ is a $\sim$\,0.9\,$\sigma$ result.

To calculate combined end station displacement factors, we follow the formalism detailed in \sref{subsec:MethodCalcs}.  We  calculate the weighted geometric mean of 1 and 1.0046, and its relative uncertainty,  using weighting based on the  \SI{0.42}{\percent} and \SI{0.32}{\percent} non-common uncertainty contributions for $X_{_X}$ and $X_{_Y}$.  The combined displacement correction factors, $C_{_{X}}$ and $C_{_{Y}}$, calculated using \eref{eq:corr_fact} together with $\chi_{_{XY}}$, $\mu_{g}$ and the non-common factors contributing to end station displacement factors, are listed in \tref{tab:C_factors} with their relative uncertainty estimates.
\begin{table}[t]
\caption{Calculated X-end and Y-end combined displacement correction factors, $C_{_X}$ and $C_{_Y}$, together with $\chi_{_{XY}}$, $\mu_g$, and the non-common factors contributing to end station displacement factor uncertainty (indented), and their uncertainties, for the LHO interferometer during the O3 observing run. For Type\,A uncertainties, the  number of measurements is noted in parentheses.}
\vspace{0.1in}
\begin{indented}
    \item[]\begin{tabular}{L{2.0cm} C{1.5cm} C{1.6cm} C{1.5cm} C{1.6cm} C{0.7cm} C{0.7cm}}
        \Xhline{4\arrayrulewidth}
        \multirow{2}{*}{\bf Param} & \multicolumn{2}{c}{LHO X-end} & \multicolumn{2}{c}{LHO Y-end} & \multirow{2}{*}{\bf Unit} & \multirow{2}{*}{\bf Type} \\
        \cline{2-3} \cline{4-5}
         &\bf  Values & $\bf u_{rel}$ \bf (\%) & \bf Values & $\bf u_{rel}$ \bf (\%)  & \\
        \Xhline{2\arrayrulewidth}
        $C_{_X}, C_{_Y}$ &  1.0029 & 0.25 & 1.0017 & 0.25  & - & C \\
        \hspace{5 mm}$\chi_{_{XY}}$  & 1.0046 & 0.01 (148) & 1.0046 & 0.01 (142) &- & A  \\ 
        \hspace{5 mm}$\mu_{g}$ & 1.003 & 0.25 & \multicolumn{2}{c}{Common with X-end}  & - & C \\  
        \hspace{10mm}$\cos \theta$   & 0.9884 & 0.03  & 0.9884 & 0.03 &- & B\\ 
        \hspace{10mm}$M$ & 39.657 & 0.01 & 39.584  & 0.01  & kg & B  \\
        \hspace{10mm}$\epsilon_{rot}$  & - & 0.41 &- & 0.31  &- & B \\  
        \hspace{10 mm}$\alpha_{_{RW}}$   & -0.7209  & 0.042 (8)  & -0.7157 &  0.014 (12) & - & A\\
        \hspace{10 mm}$\zeta_{_{W}}$   & 1636.9  &  0.002 (8)& 1637.6 & 0.002 (9) & $\rm ct/V$ & A \\
        \hspace{10mm}$\eta_{_{R}}$ & 0.9942 & 0.04 & 0.9948  & 0.04 & - & C \\  
        \hspace{10mm}$\xi_{_{EL}}$ & 0.9986 & 0.040 & 0.9986 & 0.040 & -& C\\
        \Xhline{4\arrayrulewidth}
        \end{tabular}
\label{tab:C_factors}
\end{indented}
\end{table}
The combined displacement factors, $X^c_{_{X}}$ and $X^c_{_{Y}}$, calculated using~\eref{eq:disp_coeffXY} and their relative uncertainties are listed in \tref{tab:X_prime}.   The uncertainties for these combined displacement factors have been estimated by summing the correction factor uncertainties, 0.25\,\%, in quadrature with the uncertainty of 0.32\,\% from factors that are common to both end stations, $\rho_{_{G}}$, $\alpha_{_{WG}}$ and $\xi_{_{LN}}$.  The resulting overall uncertainties of 0.41\,\% for the combined displacement factors are smaller than those for the displacement factors for each end station, 0.53\,\% for $X_{_X}$  and 0.45\,\% for $X_{_Y}$.  This reduction results from combining the calibrations from both end stations using the measured X/Y comparison factor, $\chi_{_{XY}}$.  This lowers the uncertainty contributions from sources that are not common to both end stations from 0.42\,\% for X-end and 0.32\,\% for Y-end to 0.25\,\% for $C_{_{X}}$ and $C_{_{Y}}$.
\begin{table}[t]
\caption{Measured {\em combined} displacement factors, $X^{c}$,  together with contributing factors (indented) and uncertainties, for the LHO interferometer during the O3 observing run. For Type\,A uncertainties, the  number of measurements is noted in parentheses.}
\vspace{0.1in}
\begin{indented}
    \item[]\begin{tabular}{L{2.0cm} C{1.5cm} C{1.6cm} C{1.5cm} C{1.6cm} C{0.7cm} C{0.7cm}}
        \Xhline{4\arrayrulewidth}
        \multirow{2}{*}{\bf Param} & \multicolumn{2}{c}{LHO X-end} & \multicolumn{2}{c}{LHO Y-end} & \multirow{2}{*}{\bf Unit} & \multirow{2}{*}{\bf Type} \\
        \cline{2-3} \cline{4-5}
         &\bf  Values & $\bf u_{rel}$ \bf (\%) & \bf Values & $\bf u_{rel}$ \bf (\%)  & \\
        \Xhline{2\arrayrulewidth}
        $ X^c_{_X}, X^c_{_Y}$  & \num{1.561e-14}  & 0.41 & \num{1.581e-14} & 0.41& m/ct  &C\\
        
        \hspace{5 mm}$\rho_{_G}$ & -8.0985 & 0.315 & \multicolumn{2}{c}{Common with X-end} & V/W & C\\
        
        \hspace{5 mm}$\alpha_{_{WG}}$  & 1.1172 & 0.01 (38) & \multicolumn{2}{c}{Common with X-end} & -& A  \\
        
        \hspace{5 mm}$\xi_{_{LN}}$ & 1.0020  & 0.070 & \multicolumn{2}{c}{Common with X-end} & -& C\\
               
        \hspace{5 mm}$C_{_X},C_{_Y}$ &  1.0029 & 0.25 & 1.0017 & 0.25  & -& C \\
        \Xhline{4\arrayrulewidth}
        \end{tabular}
\label{tab:X_prime}
\end{indented}
\end{table}

\section{Summary and conclusions}
\label{sec:Conc}
The estimated overall 1\,$\sigma$ relative standard uncertainties of $0.41\,\%$ 
for both the $X^{c}_{_X}$ and $X^{c}_{_Y}$ displacement factors are significantly lower than previously reported values\,\cite{P1900127}.  The reduced uncertainty in the calibration of the GS at NIST (from 0.42\,\% to 0.32\,\%) is a key contributor to this improvement.  The updated factors are corrected for the temperature dependence of the GS and WS sensors and the ambient temperature differences between laboratories where the measurements were performed ($\xi_{_{LN}}$ and $\xi_{_{EL}}$).  They also incorporate in-chamber measurements of the ratio between the input-side and output-side 
optical efficiencies, $\beta$.  These enable apportioning the overall optical efficiency ($\eta$) measured outside 
the vacuum envelope with the WS to calculate the optical efficiency factors, $\eta_{_R}$ and 
$\eta_{_T}$. 

Unintended rotation of the ETM due to Pcal forces is one of the biggest contributors to the estimated overall calibration uncertainty.    The large displacements of the interferometer beams from the centers of the ETMs (29~mm and 22~mm) are the main cause.  The interferometers have been designed to operate with the beams located close to the center of the ETM surfaces.  However, point defects in the ETM high-reflectivity coatings\,\cite{pointDefects} have required large beam displacements to optimize interferometer performance, even while suffering the deleterious impacts of operating with mis-centered beams.  
If the coatings are improved to eliminate these defects and the interferometer beams can be moved to within a few millimeters of center, the uncertainty due to unintended rotation could be reduced by a factor of ten.  
Finally, using the interferometer to compare the X-end and Y-end Pcal displacement factors (by measuring $\chi_{_{XY}}$) has enabled generating {\em combined} displacement factors, $X^c_{_X}$ and $X^c_{_Y}$, with improved accuracy and lower estimated uncertainty.

Realizing the important role laser power sensor calibration plays in the scientific impact of gravitational wave detections, there is increased interest and activity within the national metrology institute (NMI) community to improve power sensor calibration accuracy and precision.   A comparative study executed in 2009 under the auspices of the Consultative Committee for Photometry and Radiometry (CCPR) of the Bureau International des Poids et Mesures (BIPM)\,\cite{Euromet} indicated that there might be systematic differences that are as large as a few percent between power sensor calibrations performed at various NMIs.  In March 2019 the {\em GW Metrology Workshop} was held at NIST in Boulder, CO to communicate the requirements of the GW community and investigate possibilities within the NMI community for addressing them.  A bi-lateral comparison between NIST and the Physikalisch-Technische Bundesanstalt (PTB) in Braunschweig, Germany using a LIGO working standard is currently underway. Additional bi-lateral comparisons are being considered.  Furthermore, scientists at NIST are developing a new generation of primary calibration standards using bolometers and nanotube absorbers\,\,\cite{SPIE} that are expected to have significantly lower uncertainties than the standards currently being used.  These sensors may prove suitable for locating inside the vacuum envelope where several sources of uncertainty would be mitigated, providing laser power calibration directly traceable to SI units in-situ\,\cite{lehmanNIST}.

There is another significant ongoing effort within the GW community to establish a displacement reference for the Pcals that is similar to what is provided by the NIST calibrations of the GS.  It involves deploying rotating mass quadrupoles, {\em Newtonian} (Ncal) or {\em Gravity} (Gcal) calibrators, near, but outside the ETM vacuum chambers at the observatory end stations\,\cite{InoueGcal,EstevezNcal}.  Modulation of the position of the ETM is achieved via variation of the Newtonian gravitational field produced by the rotating quadrupole.  The possibility of using similar devices has been considered for a long time\,\cite{MatoneGcal}.  Prototypes are currently installed and being commissioned at the both the Virgo and the LHO observatories.

Establishing continuous, on-line displacement fiducials with sub-percent accuracy is a key element of interferometer calibration.  But achieving sub-percent calibration of the interferometer output signals, over the entire sensitive frequency band, and over long observing runs during which changes to improve interferometer performance may have been implemented, is a daunting hurdle.  To better characterize the interferometer response function, swept-sine measurements are performed regularly using the Pcal systems.  These, together with tracking of time-varying interferometer parameters using continuously-injected Pcal modulations\,\cite{DarkhanTDPs}, have contributed to dramatically improved overall interferometer calibration uncertainties for the LIGO detectors.  In some frequency bands, the estimated systematic errors during the recently-completed O3 observing run are below 2\,\%\,\cite{LilliO3systematic}. 

Currently, interferometer calibration uncertainty does not limit the extraction of source parameters from detected GW signals. However, as the SNRs of GW signals increase, optimally extracting the encoded information they carry  will require reducing interferometer calibration uncertainty.   The improvements in the accuracy of the Pcal fiducial displacements described herein should enable realizing the sub-percent interferometer calibration accuracy required to continue extending the scientific reach of the global network of GW detectors.

\section{Acknowledgements}
The authors gratefully acknowledge the assistance of X.~Chen, D.~Estevez and L.~Datrier with lab measurements and fruitful discussions with M.~Spidell, J.~Lehman, and M.~Stephens at NIST, S.~K\"{u}ck at PTB,  and with J.~Kissel, E.~Goetz, G.~Mendell, S.~Banagiri and the entire Calibration Working Group of the LIGO Scientific Collaboration (LSC).  Ongoing collaboration with the Virgo and KAGRA calibration teams is also gratefully acknowledged.

DB, SK, VB, and EP acknowledge the LSC Fellows program for supporting their research at LHO.  DB and SK are supported by NSF award PHY-1921006. EP is supported by ARC CE170100004.
LIGO was constructed by the California Institute of Technology and Massachusetts Institute of Technology with funding from the National Science Foundation, and operates under cooperative agreement PHY-1764464. Advanced LIGO was built under award PHY-0823459. 

We would like to thank all of the essential workers who put their health at risk during the COVID-19 pandemic, without whom we would not have been able to complete this work.
 
This paper carries LIGO Document Number LIGO-P2000113.

\newpage
 \section*{Appendix A}
\begin{table}[h]
\caption{List of symbols and descriptions of the parameters they represent.}
\vspace{0.1in}
\begin{indented}
    \item[]\begin{tabular}{L{1.3cm} L{8.8cm} C{1.2cm}}
        \Xhline{4\arrayrulewidth}
        \bf Symbol & \bf Parameter (units) & \bf Section  \\
        \Xhline{2\arrayrulewidth}
        $\theta$ & Angle of incidence of the Pcal beams on the ETM & \ref{subsec:IntroForces} \\
        $\epsilon_{rot}$ & Relative uncertainty due to unintended ETM rotation & \ref{subsec:IntroForces} \\
        $\vec{a}$ & Pcal center of force offset vector (m)& \ref{subsec:IntroForces} \\
        $\vec{b}$ & Interferometer beam position offset vector (m)& \ref{subsec:IntroForces} \\
        $M$ & Mass of end test mass (kg)& \ref{subsec:IntroForces} \\
        $I_p, I_y$ & Moments of inertia in pitch and yaw direction & \ref{subsec:IntroForces}\\
        $\rho_{_G}$ & GS responsivity  (V/W)   & \ref{subsec:MethodSensorcal} \\
        $\alpha_{_{WG}}$ & WS to GS responsivity ratio & \ref{subsec:MethodSensorcal}\\
        $\alpha_{_{RW}}$ & Rx to WS responsivity ratio & \ref{subsec:MethodSensorcal}\\
        $\zeta_{_W}$ & ADC conversion correction factor (ct/V) & \ref{subsec:MethodSensorcal} \\
        $\xi_{_{LN}}$ & LSB lab.\ to NIST lab.\ temperature correction factor & \ref{subsec:MethodSensorcal} \\
        $\kappa_{_G}$ & GS relative resposivity temperature coefficient (1/K) & \ref{subsec:MethodSensorcal} \\
        $\kappa_{_W}$ & WS relative resposivity temperature coefficient (1/K)  & \ref{subsec:MethodSensorcal} \\
        $\Delta T_{_{LN}}$ & LHO lab.\ and  NIST lab.\ temperature difference (K) & \ref{subsec:MethodSensorcal} \\
        $\xi_{_{EL}}$ & End station to NIST lab.\ temperature correction factor  & \ref{subsec:MethodSensorcal} \\
        $\Delta T_{_{EL}}$ & End Station and LHO lab.temperature difference (K) & \ref{subsec:MethodSensorcal} \\
        $\rho_{_R}$ & Rx power sensor responsivity (ct/W) & \ref{subsec:MethodSensorcal} \\
        $\eta$ & Overall optical efficiency & \ref{subsec:MethodSensorcal} \\
        $\beta$ & Transmitter-side to receiver-side optical efficiency ratio & \ref{subsec:MethodSensorcal} \\
        $\eta_{_R}$ & Receiver-side optical efficiency factor &  \ref{subsec:MethodSensorcal} \\
        $\eta_{_T}$ & Transmitter-side optical efficiency factor &  \ref{subsec:MethodSensorcal} \\
        $\chi_{_{XY}}$ &  X/Y calibration comparison factor & \ref{subsec:MethodCalcs} \\
        $\mu_{g}$ & Geometric mean of 1 and $\chi_{_{XY}}$ & \ref{subsec:MethodCalcs} \\
        $C_{_X}, C_{_Y}$ & X-end, Y-end combined  displacement  correction  factors & \ref{subsec:MethodCalcs} \\
        $X_{_{X}}, X_{_{Y}} $ & End station ETM displacement factors (m/ct)& \ref{subsec:MethodCalcs} \\
        $X^c_{_X}, X^c_{_Y}$ & Combined end station ETM displacement factors (m/ct)& \ref{subsec:MethodCalcs} \\
        $\phi$ & Angle between $\vec{a}$ and $\vec{b}$ &  \ref{subsec:ResultsForcecoeff}\\
        \Xhline{4\arrayrulewidth}
    \end{tabular}
\label{tab:allParams}
\end{indented}        
\end{table}

\section*{References}
\bibliographystyle{iopart-num-long.bst}
\bibliography{pcal}
\end{document}